\documentclass[10pt,aps,prl,twocolumn,superscriptaddress]{revtex4-1}
\pdfoutput=1 
\usepackage{amsfonts}
\usepackage{mathrsfs}
\usepackage{amsmath}
\usepackage{amssymb}
\usepackage{fancyhdr}
\usepackage{graphicx}
\usepackage{xspace}
\usepackage{rotating}
\usepackage[normalem]{ulem}
\usepackage{braket}
\usepackage{verbatim}
\usepackage{xcolor}
\usepackage[utf8]{inputenc}
\usepackage[medium]{titlesec}
\usepackage{bm}
\usepackage[normalem]{ulem}
\usepackage{extarrows}
\usepackage{slashed}
\usepackage{isodateo}
\usepackage{graphicx}
\usepackage{xcolor}
\usepackage[bookmarksnumbered=true,bookmarksopen=true]{hyperref}
\usepackage[hmargin=.7in,vmargin=1.1in]{geometry}
\usepackage{indentfirst}
\usepackage{cancel}

\graphicspath{{./}{./Figs/}{./figs/}{../Figs/}{./fig/}}

\usepackage{amsfonts}
\usepackage{mathtools}
\usepackage{pifont}
\usepackage{mathrsfs}
\usepackage{amsmath}
\usepackage{amssymb}
\usepackage{framed}

\newcommand{\benum}{\begin{enumerate}}
\newcommand{\eenum}{\end{enumerate}}
\newcommand{\bi}{\begin{itemize}}
\newcommand{\ei}{\end{itemize}}
\newcommand{\Mp}{M_{\rm pl}}

\newcommand{\be}{\begin{equation}}
\newcommand{\ee}{\end{equation}}

\newcommand{\bea}{\begin{eqnarray}}
\newcommand{\eea}{\end{eqnarray}}
\newcommand{\beq}{\begin{eqnarray}}
\newcommand{\eeq}{\end{eqnarray}}

\newcommand{\Rmnum}[1]{\expandafter\@slowromancap\romannumeral #1@}

\renewcommand{\FR}[2]{\displaystyle\frac{\,{#1}\,}{#2}}

\def\bge{\begin{equation}}
\def\ede{\end{equation}}
\def\bga{\begin{aligned}}
\def\eda{\end{aligned}}
\def\bgp{\begin{pmatrix}}
\def\edp{\end{pmatrix}}
\def\bgs{\begin{subequations}}
\def\eds{\end{subequations}}

\def\di{{\mathrm{d}}}

\def\mb{\mathbf}

\def\pd{\partial}
\def\ld{{\mathscr{L}}}

\def\la{\langle}\def\ra{\rangle}

\def\to{\rightarrow}
\def\To{\Rightarrow}
\def\ii{\mathrm{i}}

\def\al{\alpha}
\def\be{\beta}

\def\de{\delta}
\def\ep{\epsilon}
\def\ka{\kappa}

\def\si{\sigma}

\def\Mp{M_{\text{Pl}}}

\newcommand{\ob}[1]{\mkern 2mu \overline{\mkern -2mu #1 \mkern -2mu}\mkern 2mu}
\newcommand{\wt}[1]{\mkern 2mu \widetilde{\mkern -2mu #1 \mkern -2mu}\mkern 2mu}

\newcommand{\fr}[2]{\mbox{$\frac{\,{#1}\,}{#2}$}}
\newcommand{\n}{\nonumber}
\renewcommand{\rm}{\mathrm}

\begin{document} 
\title{Probing Leptogenesis with the Cosmological Collider
}

\author{Yanou Cui}
\email[]{yanou.cui@ucr.edu}
\affiliation{Department of Physics and Astronomy, University of California, Riverside, CA 92521, USA}
\author{Zhong-Zhi Xianyu}
\email[]{zxianyu@tsinghua.edu.cn}
\affiliation{
Department of Physics, Tsinghua University, Beijing 100084, China
}
\affiliation{
Collaborative Innovation Center of Quantum Matter, Beijing, 100084, China
}

%

\begin{abstract}
Leptogenesis is generally challenging to directly test due to the very high energy scales involved. In this work we propose a new probe for leptogenesis with cosmological collider physics. With the example of a cosmological Higgs collider, we demonstrate that during inflation leptogenesis models can produce detectable primordial non-Gaussianity with distinctive oscillatory patterns that encode information about the lepton-number violating couplings, the Majorana right-hand neutrino masses, and the CP phases, which are essential to leptogenesis. 

\end{abstract}
\maketitle
\textbf{Introduction.} 
Leptogenesis is an attractive mechanism explaining the origin of the baryon asymmetry in our universe, and is closely related to the seesaw mechanism for the origin of neutrino masses \cite{Fukugita:1986hr,Luty:1992un,Buchmuller:2004nz,Davidson:2008bu,Bodeker:2020ghk,Xing:2020ald}. 
In the simplest leptogenesis scenario, a lepton asymmetry is generated from the decay of a heavy right-handed (RH) neutrino $N$ which is then transferred to a baryon asymmetry via the sphaleron process.
Despite the appeals of leptogenesis, it is very challenging to directly test it, as the very high energy scales involved are generally beyond the reach of terrestrial experiments, in contrast to some of the low scale baryogenesis models (E.g. \cite{Cohen:1990it, Cirigliano:2006dg, Morrissey:2012db, Cui:2011ab,Cui:2012jh,Cui:2014twa,Nelson:2019fln,Cui:2020dly,Beniwal:2017eik,Chala:2016ykx,Cui:2015eba,Cui:2016rqt,Croon:2019ugf,Co:2019wyp,Elor:2020tkc}). For instance, with hierarchical masses of $N_i$ ($i=1,2,3$), a lower bound exists for the lightest $N$'s mass: $M_1\gtrsim 10^9$~GeV \cite{Davidson:2002qv}.

In this \textit{Letter} we demonstrate that, thanks to the high energies provided by the cosmic inflation, leptogenesis may leave distinct and observable signatures in primordial non-Gaussianity (NG), inspired by the Cosmological Collider (CC) physics. In particular, the essential Sakharov conditions for leptogenesis such as $L$-number violation and CP violation, as well as the heavy RH neutrino masses, can be probed individually as they impact the CC signals in different ways. Our finding also leads to new types of fermionic CC signatures that are not yet considered in the literature. 

CC physics has been recently developed as a method to probe heavy particles during cosmic inflation \cite{Arkani-Hamed:2015bza,Chen:2016nrs,Lee:2016vti,Chen:2016uwp,Chen:2016hrz,An:2017hlx,Kumar:2017ecc,Chen:2017ryl,Chen:2018xck,Wu:2018lmx,Li:2019ves,Lu:2019tjj,Liu:2019fag,Hook:2019zxa,Hook:2019vcn,Kumar:2019ebj,Alexander:2019vtb,Wang:2019gbi,Wang:2019gok,Wang:2020uic,Li:2020xwr,Wang:2020ioa,Fan:2020xgh,Aoki:2020zbj,Bodas:2020yho,Maru:2021ezc,Lu:2021wxu,Wang:2021qez,Kim:2021ida}. The energy scale of inflation, characterized by the nearly constant Hubble parameter $H$, can be up to $O(10^{13})$~GeV in single-field slow-roll models. Different versions of CC exist. The original version is an inflaton collider: by coupling to the inflaton, heavy particles with masses up to $O(H)$ imprint the $n$-point ($n\geq3$) correlators of the curvature perturbation $\zeta$. An alternative version, called the Cosmological Higgs Collider (CHC) \cite{Lu:2019tjj}, exploited the possibility that an appreciable component of $\zeta$ is originated from the quantum fluctuations of the Standard Model (SM) Higgs field  $\mb H$ during inflation, through the modulated reheating process \cite{Dvali:2003em, Kofman:2003nx, Suyama:2007bg, Ichikawa:2008ne}. Compared with the original CC, CHC has the advantage of yielding large and detectable NG with less free parameters. With these considerations, along with the fact that the SM Higgs directly participates in leptogenesis via the Yukawa coupling, in this work we will focus on signals with the CHC. 

\textbf{Cosmological Higgs Collider.} 
Quantum fluctuation during inflation generates a space dependent VEV $ v\sim H$ for the Higgs field. In the program of CHC, this VEV controls the inflaton decay rate and by doing so imprints its spatial dependence on the fluctuations of the local time of reheating. Consequently, the Higgs fluctuations $\delta h$ can source at least a fraction of primordial curvature perturbation $\zeta=\zeta_0+\zeta_h$, where $\zeta_h$ is the fluctuation from Higgs modulated reheating and $\zeta_0$ denotes other contributions such as from the inflaton fluctuations. Then we can write $\zeta_h=(2\pi P_\zeta^{1/2}/H)R_h \de h$ at the linear level, where $R_h=\zeta_h/\zeta$ and $P_\zeta\sim 2\times 10^{-9}$ is the measured scalar power spectrum. Thus the 3-point correlator (bispectrum) of $\zeta_h$ is related to the 3-point correlator of $\de h$ as
\bge
\label{3ptfunction}
  \la \zeta_{h,\mb{k}_1}\zeta_{h,\mb{k}_2}\zeta_{h,\mb{k}_3}\ra=\Big(\FR{2\pi P_\zeta^{1/2}R_h}{H}\Big)^3\la \de h_{\mb{k}_1}\de h_{\mb{k}_2}\de h_{\mb{k}_3}\ra.
\ede
In the original proposal \cite{Lu:2019tjj} the fraction $R_h$ is roughly 10\%, but this is model dependent. The upshot is that $\zeta_h$, while being a small fraction of $\zeta$, can contribute to most of the NG. In this \emph{Letter} we treat $R_h$ as a free parameter, $0<R_h<1$. Furthermore, the working of CHC requires that the Higgs potential is stabilized and that the Higgs is lighter than $H$ during the inflation. While this might call for new physics beyond SM, we note that the signal considered below only depends on the neutrino properties and is essentially independent of the new physics that stabilized the Higgs potential. 

\begin{figure}
\parbox{0.26\textwidth}{\vspace{-4mm}\includegraphics[width=0.26\textwidth]{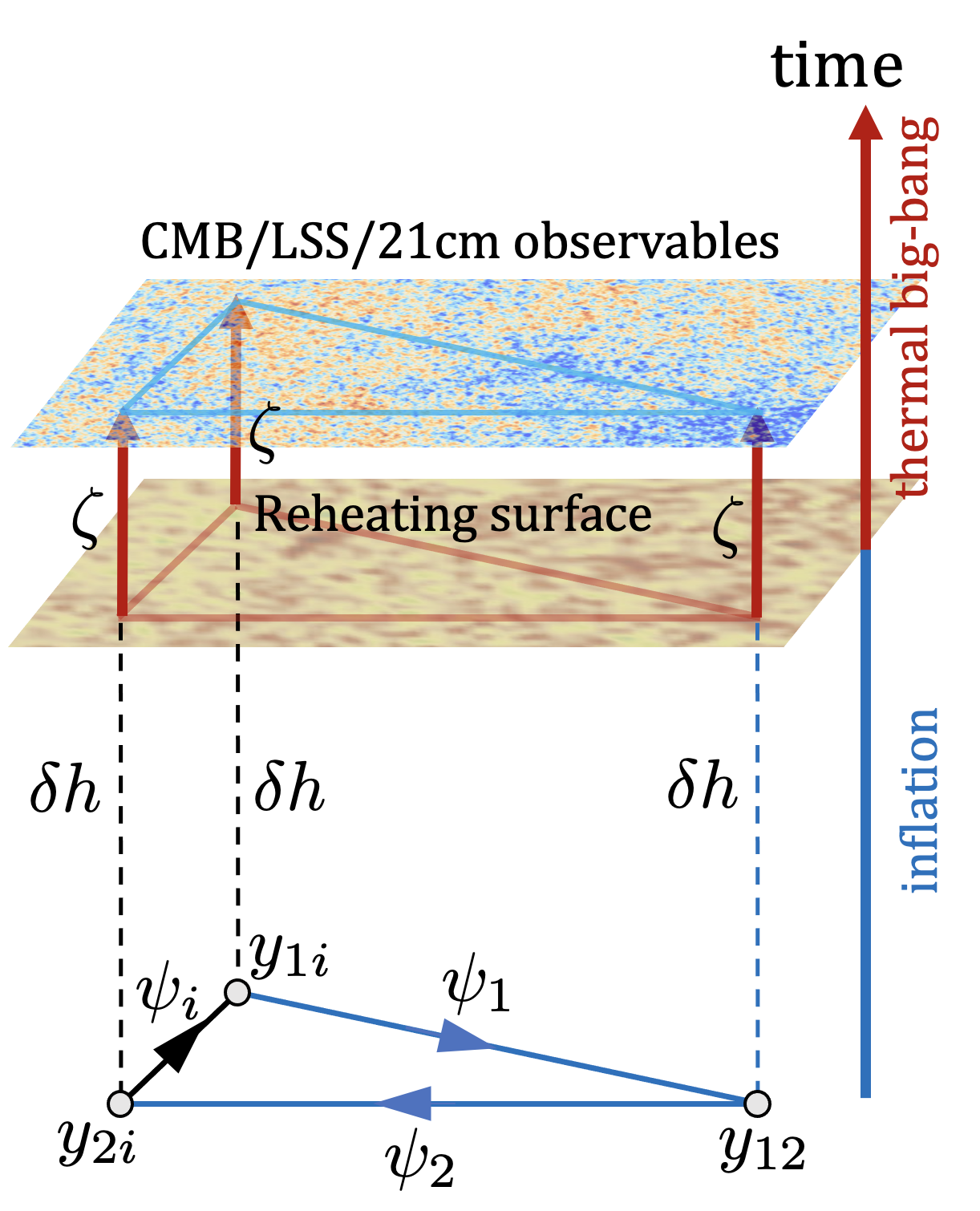}\vspace{-2mm}}
$\To$~
\parbox{0.14\textwidth}{\includegraphics[width=0.14\textwidth]{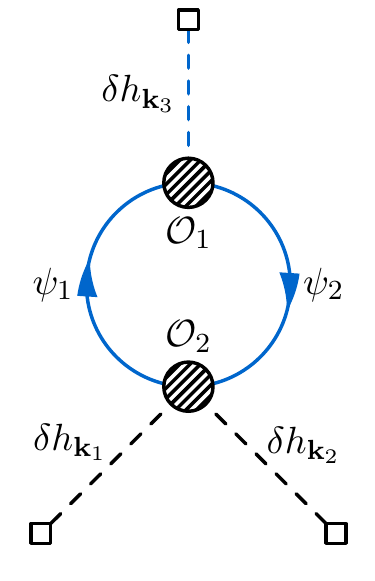}}
\caption{1-loop process contributing to SM and Majorana RH neutrino signals at the CHC ($i=1,2$).}
\label{fig_1loopdiag}
\end{figure}

Eq.~(\ref{3ptfunction}) shows that the primordial bispectrum in CHC can reveal information about Higgs interactions during inflation. In particular, particles with masses $m\sim H$ may leave a signal in the bispectrum if they couple to the SM Higgs. In this \emph{Letter} we are interested in the signals mediated by neutrinos represented by the 1-loop diagram on the left side of Fig.~\ref{fig_1loopdiag} (we will explain the right diagram in the next section). 

The signal shows up for squeezed bispectrum where $k_1\simeq k_2\gg k_3$ ($k_i\equiv|\mb k_i|$, $i=1,2,3$), as a characteristic oscillatory function of the momentum ratio $\varrho\equiv k_3/k_1$. It is useful to express the signal in terms of a dimensionless and (nearly) scale-invariant shape function $\mathcal{S}(k_1,k_2,k_3)\equiv (k_1k_2k_3)^2/(4\pi^2 P_\zeta)^2\la \zeta_{h,\mb{k}_1}\zeta_{h,\mb{k}_2}\zeta_{h,\mb{k}_3}\ra'$ ($'$ symbol means the momentum-conserving $\delta$-function removed), as (see \cite{Wang:2021qez} for more details):
\begin{align}
\label{squeezedshape}
  \lim_{\varrho\to 0}\mathcal{S}(\varrho)=f_\text{NL}^\text{(signal)}\varrho^\al \sin\big(\omega\log\varrho+\vartheta\big),
\end{align}
where $f_\text{NL}^\text{(signal)}$ is the amplitude of the signal. The dimensionless quantities  ($\al, \omega, \vartheta)$ measure the scaling behavior, the frequency, and the phase of the oscillatory signal. They are in principle measurable and also calculable for specific processes.  It is known that the signal with a Dirac fermion of mass $m$ in a 1-loop diagram has $\al=3$ and $\omega=2m/H$ \cite{Lu:2019tjj}. A key result of this \textit{Letter} is that the Majorana mass of the RH neutrino and CP phases innate to the leptogenesis mechanism can generate new patterns in the signal. In particular, the CHC signal with a Majorana neutrino gives $\al=2$ instead of 3, and the CP phases give rise to more than one value of $\omega$, and the signal would be superpositions of different oscillation patterns.

\textbf{Leptogenesis, Neutrino masses and CP phases during inflation.} 
Leptogenesis mechanism meets the Sakharov conditions in the following ways: 1) lepton number violation via the Majorana masses of the RH neutrinos $N$'s; 2) CP violation via the CP phases in the Yukawa couplings through the tree-loop interference in $N$ decay; and 3) out-of-equilibrium via the decay process \cite{Sakharov:1967dj, Fukugita:1986hr}. 

A key observation is that, with a Majorana mass term, the mass matrix and the Higgs couplings cannot be simultaneously diagonalized. This results in a new type of Yukawa coupling mixing different mass eigenstates with distinct CHC signals. The following derivations closely follow the standard neutrino seesaw model, with the key difference that $\mb H$'s VEV is typically around the inflationary Hubble scale $H$, and consequently the neutrino mass/mixing pattern is very different in the inflation and post-inflation eras.

One generation suffices to demonstrate how the mixed coupling arises. Consider the usual Type I seesaw with one generation of $N$ and SM lepton doublet $L=(\nu,e^-)^T$. In CHC the Higgs gets a nonzero VEV $v\sim H$ during inflation due to quantum fluctuations. Here the inflationary Hubble parameter $H$ is required to be higher than electroweak (EW) scale for the original CHC mechanism \cite{Lu:2019tjj} to work.
Parametrizing the Higgs as $\mb H=(0,(v+h)/\sqrt{2})^T$ and focusing on neutrinos only, we have
\begin{align}
  \Delta\ld=&~\nu^\dag\ii\ob\si^\mu\pd_\mu \nu+ N^\dag\ii\ob\si^\mu\pd_\mu N\n\\
  &~+\Big[m_D\big(1+\FR{h}{v}\big)\nu N-\FR{1}{2}m_N NN+\text{c.c.}\Big],
\end{align}
where $m_D\equiv yv/\sqrt{2}$. The $U(2)$ symmetry of the kinetic term allows us to choose $m_D$ and $m_N$ to be real without loss of generality, and also to rotate to mass eigenstates $\psi_\pm$ with   mass eigenvalues $m_\pm=\fr{1}{2}(m_N\pm\sqrt{m_N^2+4m_D^2})$. In the post-inflationary epoch, Higgs resides in its true vacuum such that $m_D\ll m_N$, and thus $m_-\ll m_+$, realizing the seesaw mechanism after EWPT. But during inflation $m_D\sim m_N\sim H$ is likely, therefore $\psi_\pm$ could have comparable masses. 

Although these are mostly familiar results, the important point is that the Higgs coupling matrix cannot be simultaneously diagonalized by the above rotations. Rather, it takes the following form after the rotation:
\begin{align}
  \ld\supset 
  \FR{m_D h}{v\sqrt{m_N^2+4m_D^2}}\Big[m_D(\psi_-^2-\psi_+^2)+m_N\psi_-\psi_+\Big],
\end{align}
with a nonzero trilinear coupling that mixes $\psi_\pm$ when $m_N\neq0$. As we will see, this mixed Yukawa coupling in the mass eigenstates can give rise to a unique signal in 3-point function of Higgs, and thus can  be viewed as a signature of Majorana neutrino mass. 

Realistic leptogenesis model typically involves 3 generations of RH $N$'s. With nonzero Majorana mass terms, the above result of Yukawa coupling to mixed mass eigenstates still apply, but with the new feature that these couplings  generally contain irremovable CP phases. As we will see below, these phases introduce yet another new pattern in the CHC signal. 

\textbf{Cosmological (Higgs) Collider Signals of Leptogenesis.}
A key result from the above discussion is that the Yukawa couplings mix different mass eigenstates and carry nonzero complex phases. Now we demonstrate their distinctive physical consequences on the CHC observables. 

The central object is the 3-point correlator of the Higgs fluctuation, namely the left diagram in Fig.~\ref{fig_1loopdiag}. 
The CHC signal appears when the blue lines carry momenta much smaller than the black lines, i.e.,  in the squeezed limit. In this configuration it proves advantageous to take an EFT limit for the bottom black line in the triangle, which effectively shrinks the line into a point vertex \cite{Tong:2021wai}, as shown in the right diagram of Fig.~\ref{fig_1loopdiag}. Then it is straightforward to use the diagrammatic rule \cite{Chen:2017ryl} to write down an expression for this process:
\begin{align}
\label{3pt}
  &\la \de h_{\mb{k}_1}\de h_{\mb{k}_2}\de h_{\mb{k}_3}\ra'=\sum_{\mathsf{a,b}=\pm}\mathsf{ab}\int_{-\infty}^0\di\tau_1\di\tau_2 a^4(\tau_1)a^4(\tau_2)\n\\
  &~~\times G_\mathsf{a}(k_1,\tau_1)G_\mathsf{a}(k_2,\tau_1)G_\mathsf{b}(k_3,\tau_2)\mathcal{I}(k_3;\tau_1,\tau_2),
\end{align} 
where $\tau$ is conformal time, $a=-1/(H\tau)$ is the scale factor, $\mathsf{a},\mathsf{b}=\pm$ are Schwinger-Keldysh indices, $G_\mathsf{a}$ is the boundary-to-bulk $\de h$ propagator and $\mathcal{I}$ is the fermion 1-loop integral including couplings: 
\bge
  \mathcal{I}(k;\tau_1,\tau_2)=\int\di^3 X\, e^{-\ii\mb k\cdot\mb X} \la\mathcal{O}_1(\tau_1,\mb X)\mathcal{O}_2(\tau_2,\mb 0)\ra. 
\ede
Now we explain the effective operators $\mathcal{O}_{1,2}$.  We can write the most general Yukawa coupling mixing a pair of mass eigenstate as
\bge
\label{o1real}
  \Delta\ld=\de h\mathcal{O}_1,~~\mathcal{O}_1\equiv y_{12}(e^{\ii\varphi_{12}}\psi_1\psi_2+\text{c.c.}),
\ede
where $\psi_{1,2}$ are Weyl spinors and mass eigenstates with masses $m_{1,2}$. Generally, $m_1\neq m_2$ ($m_1=m_2$) with (without) the presence of Majorana mass in the original Lagrangian. We set $y_{12}$  real and positive and isolate the CP phase $\varphi_{12}$. $\mathcal{O}_1$ accounts for the upper vertex in the right diagram in Fig.~\ref{fig_1loopdiag}. The diagonal case $\mathcal{O}_1=y_{ii}\psi_i\psi_i+$ c.c. $(i=1,2)$ can also contribute to the signal, but this is not mass mixing and the signal would be identical to the case of ``Dirac'' to be discussed below.
When the diagonal case contribution is present and sizable, the new signal in this work can be extracted by dedicated template fitting.

The lower vertex in the right diagram in Fig.~\ref{fig_1loopdiag} is an effective vertex with the ``local'' neutrino integrated out as explained above. This effective coupling takes the form
\bge
\Delta\ld=\FR{1}{2}(\de h)^2\mathcal{O}_2,~~
  \mathcal{O}_2\equiv \FR{1}{\Lambda}(e^{\ii\varphi_5}\psi_1\psi_2+\text{c.c.}),
\ede
where we denote the absolute value of effective coupling $y_{12}(y_{11}/m_1+y_{22}/m_2)$ by an effective cutoff $1/\Lambda$ and its phase by $\varphi_5$.

We take Wick contraction and the spinor trace to get:
\begin{align}
\label{O1O2}
  & \la \mathcal{O}_1(x)\mathcal{O}_2(y)\ra
  =-\FR{4y_{12}}{\Lambda}\Big[\cos(\varphi_{12}+\varphi_5) g_{m_1}(x,y)g_{m_2}(x,y)\n\\
  &~+\cos(\varphi_{12}-\varphi_5) f_{m_1}(x,y)f_{m_2}(x,y)\Big].
\end{align}
Here $g_m$ and $f_m$ are functions of mass $m$ and positions $x=(\tau_1,\mb X)$ and $y=(\tau_2,\mb 0)$, given explicitly in \cite{Chen:2018xck}. The minus sign comes from the anticommutativity of spinors. 

 The oscillatory signal in CHC bispectrum can be found by expanding $f_m$ and $g_m$ in the late time limit $\tau_{1,2}\to 0$: 
\begin{align}
  &\ii f_m(x,y)= 2\text{Re}\bigg\{\FR{\Gamma(2-\ii \wt m)\Gamma(\fr{1}{2}+\ii \wt m)}{4\pi^{5/2}}\Big(\FR{\tau_1\tau_2}{X^2}\Big)^{3/2-\ii \wt m}\n\\
  &\times\bigg[1+\FR{\big(3-4 \wt m(2\ii + \wt m)\big)(\tau_1^2+\tau_2^2)-6\tau_1\tau_2}{2(1-2\ii\wt m)X^2}\bigg]\bigg\},\\
  &g_m(x,y)=2\text{Re}\bigg\{\FR{\Gamma(2-\ii \wt m)\Gamma(\fr{1}{2}+\ii \wt m)}{4\pi^{5/2}}\Big(\FR{\tau_1\tau_2}{X^2}\Big)^{3/2-\ii \wt m}\n\\
  &~\times\bigg[1+\FR{\big(3-4\wt m(2\ii +\wt m)\big)(\tau_1^2+\tau_2^2)+6\tau_1\tau_2}{2(1-2\ii\wt m)X^2}\bigg]\bigg\}.
\end{align} 
Here and later, any mass with a tilde denotes its dimensionless value in unit of $H$, namely $\wt m=m/H$, and $X\equiv|\mb X|$.
We see that $\ii f_m$ and $g_m$ are identical at the leading order and differ at subleading orders. This has important consequences:

1. For pure Dirac mass, the Yukawa coupling is diagonalizable with real eigenvalues. Therefore, in (\ref{O1O2}), $\la\mathcal{O}_1\mathcal{O}_2\ra$ is proportional to $f_{m_i}^2+g_{m_i}^2$ $(i=1,2)$. This combination is zero at the leading order in $\tau\to 0$ limit, and shows up only at subleading orders in $\tau$. In the in-in correlator, the power-law dependence on $\tau$ gets translated to a scaling in the $k$-ratio. Consequently, the signal with only Dirac mass decays faster than naively expected in the squeezed limit. This result is known \cite{Chen:2018xck,Lu:2019tjj} and is compatible with the scenario of Dirac neutrino/Dirac leptogenesis.

2. If there is a Majorana-mass-induced mixed Yukawa coupling but without CP phases, the correlator (\ref{O1O2}) 
would be proportional to $f_{m_1}f_{m_2}+g_{m_1}g_{m_2}$  with $m_1\neq m_2$. Consequently, a piece proportional to $(\tau_1\tau_2)^{\pm\ii(\wt m_1-\wt m_2)}$ is not canceled out at the leading order. We then expect to see an oscillating signal at the leading order of squeezeness with a \emph{single} frequency given by $\wt m_1-\wt m_2$. The other combination $\wt m_1+\wt m_2$ gets cancelled for the same reason as the Dirac case.

3. If, in addition, the mixed Yukawa coupling contains irremovable CP phases (as in the realistic leptogenesis models), the signal would be proportional to $\cos(\varphi_{12}+\varphi_5) f_{m_1}f_{m_2}+\cos(\varphi_{12}-\varphi_5) g_{m_1}g_{m_2}$. Then, both $(\tau_1\tau_2)^{\pm\ii(\wt m_1-\wt m_2)}$ and $(\tau_1\tau_2)^{\pm\ii(\wt m_1+\wt m_2)}$ show up at the leading order for generic $\varphi_{12}$ and $\varphi_5$. Therefore, this new feature would reveal itself as two distinct sets of oscillation modes in the CHC signature at the leading order of the squeezeness. 
\begin{figure}
\centering
\includegraphics[width=0.43\textwidth]{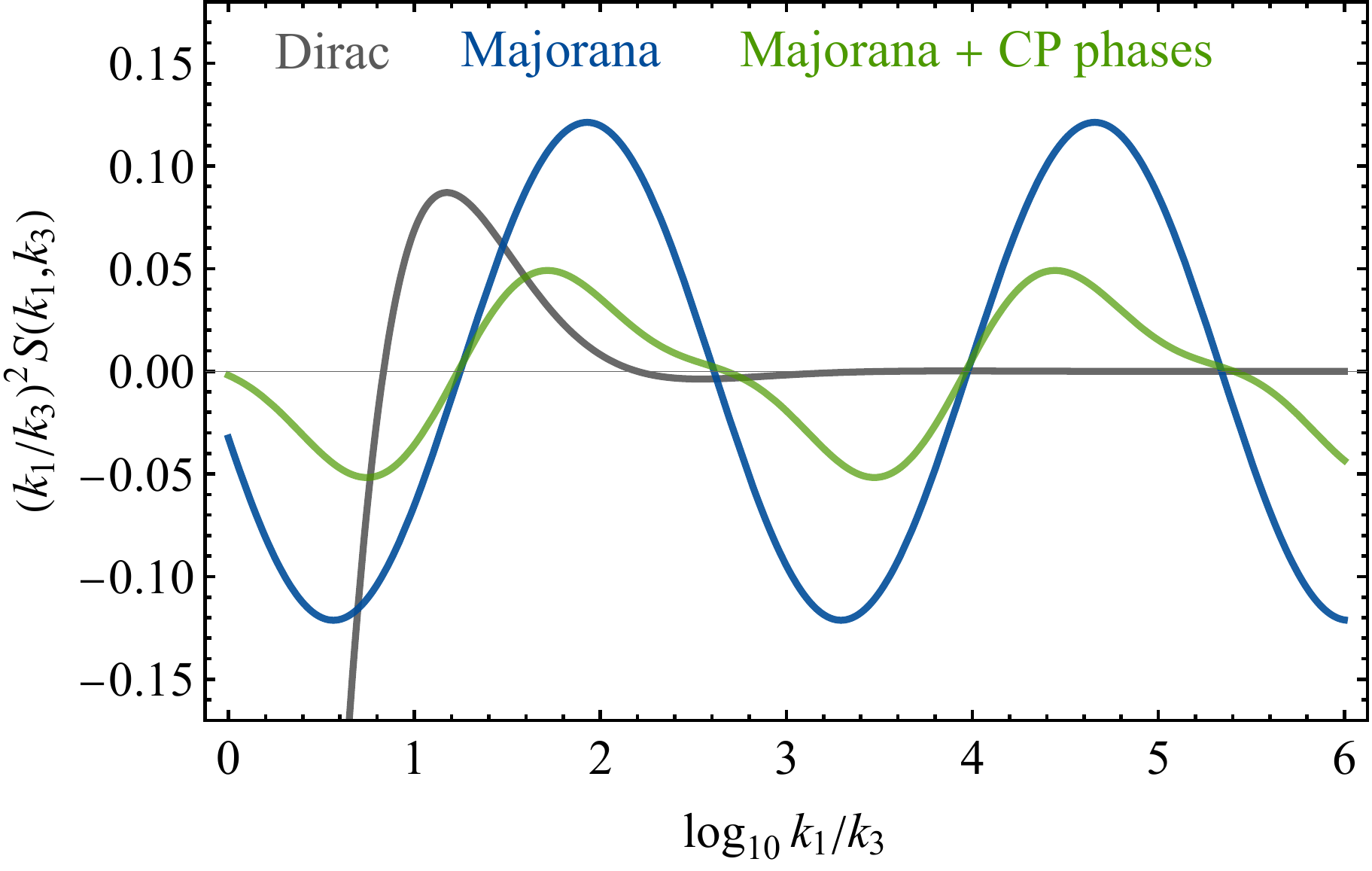}
\caption{The oscillatory shape function of the primordial bispectrum (\ref{signalshape}) as a function of $k_1/k_3$ for the three cases we consider in the text. We take $m_D=0.5H$ for ``Dirac'', and take $m_1=0.5H$ and $m_2=1.5H$ for both ``Majorana'' and ``Majorana$+$CP.'' We take $\varphi_{12}=\pi/4$ and $\varphi_5=-\pi/3$ for ``Majorana$+$CP.'' For all cases we take $R_h^3y_{12}H/\Lambda=1$.}
\label{fig_signal}
\end{figure}

It is straightforward to compute the 3-point correlator of $\delta h$ and the shape function in (\ref{squeezedshape}). The result for non-Dirac case is: (See \cite{Chen:2018xck,Lu:2019tjj} for results of pure Dirac mass.)
\begin{align}
\label{signalshape}
    &\lim_{\varrho\to 0}\mathcal{S}(\varrho)= 2R_h^3 P_\zeta^{-1/2}\FR{y_{12}}{\Lambda}\n\\
    &\times\big\{\cos\varphi_5\cos\varphi_{12}\text{Re}\big[\mathcal{C}(\wt m_1,\wt m_2)\varrho^{2+\ii(\wt m_1-\wt m_2)}\big]\n\\
&+\sin\varphi_5\sin\varphi_{12}\text{Re}\big[\mathcal{C}(\wt m_1,-\wt m_2)\varrho^{2+\ii(\wt m_1+\wt m_2)}\big]\big\},
\end{align}
where
\begin{align}
\label{Cfactor}
  &\mathcal{C}(\wt m_1,\wt m_2)=\FR{\ii}{2^{1+\ii\wt m_{12}}\pi^5}\FR{1}{\wt m_{12}^2}(1+\FR{\ii\wt m_{12}}{4})(1-\cosh\wt m_{12}\pi)\n\\
  &\times\sinh(\wt m_{12}\pi)\Gamma(\fr{1}{2}-\ii\wt m_1)\Gamma(\fr{1}{2}+\ii\wt m_2)\Gamma(2+\ii\wt m_1)\n\\
  &\times\Gamma(2-\ii\wt m_2)\Gamma^2(2+\ii\wt m_{12})\Gamma(-4-2\ii\wt m_{12}),
\end{align}
with $\wt m_{12}\equiv \wt m_1-\wt m_2$. The phase Arg\,$\mathcal{C}(\wt m_1,\wt m_2)$ gives rise to the phase of the oscillatory signal, namely $\vartheta$ in (\ref{squeezedshape}), which can be used to distinguish our signal from other oscillatory loop signals \cite{Qin:2022lva}.

Fig.~\ref{fig_signal} illustrates the signal shape function for the three cases outlines above, with ``Dirac" known in the literature, and ``Majorana" and ``Majorana+CP phases" our new results featuring different oscillation patterns. ``Majorana+CP phases" is most relevant to a realistic leptogenesis model. The ``Majorana" case is also worth noting as it represents a class of models with mass-mixing couplings with new oscillation pattern. We only present result for one pair of mixed mass eigenstates. With three generations of neutrinos, more than one pair of mass eigenstates could contribute to the signal. However, according to (\ref{Cfactor}), the signal strength depends sensitively on the masses. Thus, in general, we expect only one pair of states contribute predominantly.

\textbf{Conditions for Baryon Asymmetry and Results.}
We numerically scan parameter space to identify the regions with observable CHC signals and successful leptogenesis.
As we can see from (\ref{signalshape}), detectable CHC signals favor large Yukawa couplings. But large Yukawa couplings poses a potential tension with leptogenesis as they generally 
lead to strong washout, which we briefly review now.

\begin{figure}\label{fig2}
\centering
\includegraphics[width=0.45\textwidth]{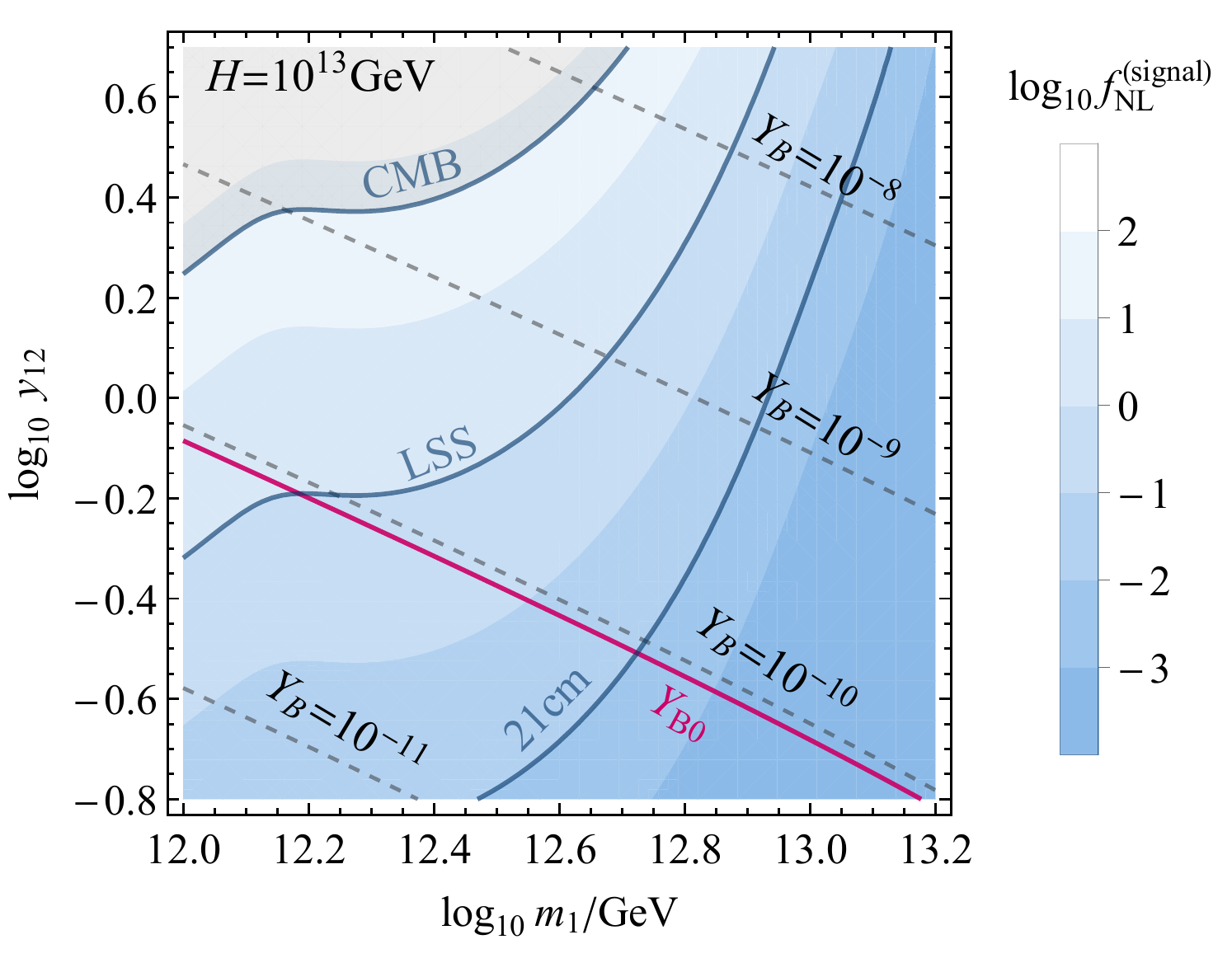}
\caption{Contour plots for the amplitude of neutrino signal at CHC $f_\text{NL}^\text{(signal)}$ and baryon density $Y_B$ predicted by the leptogenesis model on the 2D plane of Yukawa coupling $y_{12}$ and Majorana RH neutrino mass $m_1$. The solid red line indicates the observed value of baryon asymmetry, $Y_{B0}$. The solid blue lines represent experimental sensitivities to $f_{\rm NL}$.}
\label{fig_yvsm}
\end{figure}

The baryon asymmetry $Y_B$ as predicted from leptogenesis can be generally expressed as
\begin{align}
  &Y_B=\FR{c_s}{c_s-1}\ka\FR{\ep_1}{g_*},
\end{align}
where $c_s=(8N_f+4)/(22N_f+13)$ is the sphaleron conversion efficiency factor, and $c_s=28/79\simeq 0.35$ for $N_f=3$. $g_*$ is the number of relativistic degrees during leptogenesis epoch, which is $106.75$ for SM. $\ep_1$ is the asymmetry from the the decay of heavy RH neutrinos (assuming the lightest $N_1$ dominates the contribution), given by \cite{Chen:2007fv}
\bge
\label{ep1}
  \ep_1\simeq-\FR{3}{8\pi}\FR{1}{(y_\nu y_\nu^\dag)_{11}}\sum_{i=2,3}\text{Im}\Big[(y_\nu y_\nu^\dag)_{1i}^2\Big]\FR{m_1}{m_i}.
\ede
The above baryon asymmetry is subject to potential reduction by ``washout'' processes, including the inverse decay and $2\rightarrow2$ $\cancel{L}$ scattering, characterized by the washout factor $\ka$, which is the ratio of the decay rate of the RH neutrino, $\Gamma_1$, to the the Hubble scale $H$ when the temperature equals $m_1$, $r=\Gamma_1/H(T=m_1)$, i.e.,
\bge
  r=\FR{\Mp}{32\pi\times 1.7\sqrt{g_*}}\FR{(y_\nu y_\nu^\dag)_{11}}{m_1}.
\ede
$r\ll1$ is the weak washout regime, while $r\gg1$ leads to strong washout that significantly suppresses  the yield of $Y_B$.
The relation between $r$ and $\kappa$ can be obtained by solving Boltzmann equations relevant for $Y_B$ evolution \cite{Kolb:1979ui,Kolb:1990vq}. For parameters in Fig.~\ref{fig_yvsm} we are always in the moderate washout scenario ($10<r<10^6$) where the approximation $\ka\simeq 0.3/(r\log r)^{0.6}$ works well \cite{Kolb:1990vq}.

In Fig.~\ref{fig_yvsm} we show the signal size in (\ref{signalshape}) with $m_2=2m_1$, and scan over a range of mass $m_1$ and Yukawa coupling $y_{12}$ within the perturbative regime. More concretely, we take $R_h^3=1/2$, $1/\Lambda=y_{12}^2/m_1$, $\varphi_{12}=\pi/4$, $\varphi_5=-\pi/3$.
We also show the predicted baryon asymmetry $Y_B$ with its observed value today $Y_{B0}\equiv\frac{n_B-n_{\bar{B}}}{s}\simeq8.7\times10^{-11}$ from CMB and BBN related measurements~\cite{Planck:2018vyg, ParticleDataGroup:2018ovx}. In Fig.~\ref{fig_yvsm}, leptogenesis parameters realizing the observed value is shown with the solid magenta line, while contours with larger or smaller $Y_B$ are also shown to account for possible late-time dilution or the presence of other sources for baryogenesis.

 In Fig.\ \ref{fig_yvsm} we also include the reach of current CMB and forecast for future LSS/21cm observations \cite{Wang:2019gbi}. As we can see, a good range of the parameter space for viable leptogenesis lead to signals within reach of future CMB/LSS/21 cm line experiments \cite{Meerburg:2016zdz,MoradinezhadDizgah:2018ssw,Meerburg:2019qqi,Kogai:2020vzz}. Furthermore, with a CHC calibrated with known SM processes \cite{Lu:2019tjj}, it is possible to identify the neutrino signals out of SM ``backgrounds.'' 
The prospect for CHC signals is most distinct and promising when $m_1\sim H$: for $m_1\gg H$, the signal would be strongly suppressed as shown in Fig.~\ref{fig_signal}, while for $m_1\ll H$ the Majorana mass becomes subdominant to the Dirac mass term--in this case a CHC signal can be observable but restores known patterns in the literature. Furthermore, since the signal strength depends on $H$ mainly through the ratio $m_1/H$, for a fixed $m_1$, increasing (decreasing) $H$ in Fig.~\ref{fig_yvsm} amounts to shifting all shadings horizontally towards the right (left) side, while the lines remain the same.

\textbf{Discussion and Conclusion.}
In this \textit{Letter} we propose a new cosmological probe of the well-motivated leptogenesis with Majorana neutrinos, which to date is considered challenging to test directly due to the high energies involved. Specifically, we consider the scenario with SM Higgs contributing to the primordial fluctuation during inflation. Based on Cosmological (Higgs) Collider Physics, we demonstrated that this scenario can lead to observable and distinct imprints of the Higgs' Yukawa couplings to heavy RH Majorana neutrinos and SM neutrinos in the primordial bispectrum. Essential information about leptogenesis, such as the $L$-violation, heavy RH neutrino masses  (in the unit of the inflationary Hubble scale), and the CP  phases, can be extracted from delicate measurements of the primordial non-Gaussianity. With a distinct oscillatory feature, our signal is easily distinguishable from the local-shaped NG produced by the Higgs self-coupling \cite{DeSimone:2012gq}. Also, by using the phase information \cite{Qin:2022lva}, we can further select our signal out of other possible oscillatory signals contributed from addition heavy states, and thus establish the uniqueness of our signal. 
Our work presents an intriguing case of how CC physics may shed light on the profound puzzle of matter-antimatter asymmetry in our Universe, in light of the rich incoming data from CMB and LSS observations targeting primordial non-Gaussianity.

\begin{acknowledgments}
\textbf{Acknowledgements.}
We thank KITP (supported by National Science Foundation under Grant No. NSF PHY-1748958) for hospitality.
YC is supported in part by the US Department of Energy under award number DE-SC0008541.
ZX is supported in part by the National Key R\&D Program of China (2021YFC2203100), an Open Research Fund of the Key Laboratory of Particle Astrophysics and Cosmology, Ministry of Education of China, and a Tsinghua University Initiative Scientific Research Program.
\end{acknowledgments}

\bibliography{CosmoCollider}

\begin{thebibliography}{68}%
\makeatletter
\providecommand \@ifxundefined [1]{%
 \@ifx{#1\undefined}
}%
\providecommand \@ifnum [1]{%
 \ifnum #1\expandafter \@firstoftwo
 \else \expandafter \@secondoftwo
 \fi
}%
\providecommand \@ifx [1]{%
 \ifx #1\expandafter \@firstoftwo
 \else \expandafter \@secondoftwo
 \fi
}%
\providecommand \natexlab [1]{#1}%
\providecommand \enquote  [1]{``#1''}%
\providecommand \bibnamefont  [1]{#1}%
\providecommand \bibfnamefont [1]{#1}%
\providecommand \citenamefont [1]{#1}%
\providecommand \href@noop [0]{\@secondoftwo}%
\providecommand \href [0]{\begingroup \@sanitize@url \@href}%
\providecommand \@href[1]{\@@startlink{#1}\@@href}%
\providecommand \@@href[1]{\endgroup#1\@@endlink}%
\providecommand \@sanitize@url [0]{\catcode `\\12\catcode `\$12\catcode
  `\&12\catcode `\#12\catcode `\^12\catcode `\_12\catcode `\%12\relax}%
\providecommand \@@startlink[1]{}%
\providecommand \@@endlink[0]{}%
\providecommand \url  [0]{\begingroup\@sanitize@url \@url }%
\providecommand \@url [1]{\endgroup\@href {#1}{\urlprefix }}%
\providecommand \urlprefix  [0]{URL }%
\providecommand \Eprint [0]{\href }%
\providecommand \doibase [0]{http://dx.doi.org/}%
\providecommand \selectlanguage [0]{\@gobble}%
\providecommand \bibinfo  [0]{\@secondoftwo}%
\providecommand \bibfield  [0]{\@secondoftwo}%
\providecommand \translation [1]{[#1]}%
\providecommand \BibitemOpen [0]{}%
\providecommand \bibitemStop [0]{}%
\providecommand \bibitemNoStop [0]{.\EOS\space}%
\providecommand \EOS [0]{\spacefactor3000\relax}%
\providecommand \BibitemShut  [1]{\csname bibitem#1\endcsname}%
\let\auto@bib@innerbib\@empty
\bibitem [{\citenamefont {Fukugita}\ and\ \citenamefont
  {Yanagida}(1986)}]{Fukugita:1986hr}%
  \BibitemOpen
  \bibfield  {author} {\bibinfo {author} {\bibfnamefont {M.}~\bibnamefont
  {Fukugita}}\ and\ \bibinfo {author} {\bibfnamefont {T.}~\bibnamefont
  {Yanagida}},\ }\href {\doibase 10.1016/0370-2693(86)91126-3} {\bibfield
  {journal} {\bibinfo  {journal} {Phys. Lett. B}\ }\textbf {\bibinfo {volume}
  {174}},\ \bibinfo {pages} {45} (\bibinfo {year} {1986})}\BibitemShut
  {NoStop}%
\bibitem [{\citenamefont {Luty}(1992)}]{Luty:1992un}%
  \BibitemOpen
  \bibfield  {author} {\bibinfo {author} {\bibfnamefont {M.~A.}\ \bibnamefont
  {Luty}},\ }\href {\doibase 10.1103/PhysRevD.45.455} {\bibfield  {journal}
  {\bibinfo  {journal} {Phys. Rev. D}\ }\textbf {\bibinfo {volume} {45}},\
  \bibinfo {pages} {455} (\bibinfo {year} {1992})}\BibitemShut {NoStop}%
\bibitem [{\citenamefont {Buchmuller}\ \emph {et~al.}(2005)\citenamefont
  {Buchmuller}, \citenamefont {Di~Bari},\ and\ \citenamefont
  {Plumacher}}]{Buchmuller:2004nz}%
  \BibitemOpen
  \bibfield  {author} {\bibinfo {author} {\bibfnamefont {W.}~\bibnamefont
  {Buchmuller}}, \bibinfo {author} {\bibfnamefont {P.}~\bibnamefont {Di~Bari}},
  \ and\ \bibinfo {author} {\bibfnamefont {M.}~\bibnamefont {Plumacher}},\
  }\href {\doibase 10.1016/j.aop.2004.02.003} {\bibfield  {journal} {\bibinfo
  {journal} {Annals Phys.}\ }\textbf {\bibinfo {volume} {315}},\ \bibinfo
  {pages} {305} (\bibinfo {year} {2005})},\ \Eprint
  {http://arxiv.org/abs/hep-ph/0401240} {arXiv:hep-ph/0401240} \BibitemShut
  {NoStop}%
\bibitem [{\citenamefont {Davidson}\ \emph {et~al.}(2008)\citenamefont
  {Davidson}, \citenamefont {Nardi},\ and\ \citenamefont
  {Nir}}]{Davidson:2008bu}%
  \BibitemOpen
  \bibfield  {author} {\bibinfo {author} {\bibfnamefont {S.}~\bibnamefont
  {Davidson}}, \bibinfo {author} {\bibfnamefont {E.}~\bibnamefont {Nardi}}, \
  and\ \bibinfo {author} {\bibfnamefont {Y.}~\bibnamefont {Nir}},\ }\href
  {\doibase 10.1016/j.physrep.2008.06.002} {\bibfield  {journal} {\bibinfo
  {journal} {Phys. Rept.}\ }\textbf {\bibinfo {volume} {466}},\ \bibinfo
  {pages} {105} (\bibinfo {year} {2008})},\ \Eprint
  {http://arxiv.org/abs/0802.2962} {arXiv:0802.2962 [hep-ph]} \BibitemShut
  {NoStop}%
\bibitem [{\citenamefont {Bodeker}\ and\ \citenamefont
  {Buchmuller}(2021)}]{Bodeker:2020ghk}%
  \BibitemOpen
  \bibfield  {author} {\bibinfo {author} {\bibfnamefont {D.}~\bibnamefont
  {Bodeker}}\ and\ \bibinfo {author} {\bibfnamefont {W.}~\bibnamefont
  {Buchmuller}},\ }\href {\doibase 10.1103/RevModPhys.93.035004} {\bibfield
  {journal} {\bibinfo  {journal} {Rev. Mod. Phys.}\ }\textbf {\bibinfo {volume}
  {93}},\ \bibinfo {pages} {035004} (\bibinfo {year} {2021})},\ \Eprint
  {http://arxiv.org/abs/2009.07294} {arXiv:2009.07294 [hep-ph]} \BibitemShut
  {NoStop}%
\bibitem [{\citenamefont {Xing}\ and\ \citenamefont
  {Zhao}(2021)}]{Xing:2020ald}%
  \BibitemOpen
  \bibfield  {author} {\bibinfo {author} {\bibfnamefont {Z.-z.}\ \bibnamefont
  {Xing}}\ and\ \bibinfo {author} {\bibfnamefont {Z.-h.}\ \bibnamefont
  {Zhao}},\ }\href {\doibase 10.1088/1361-6633/abf086} {\bibfield  {journal}
  {\bibinfo  {journal} {Rept. Prog. Phys.}\ }\textbf {\bibinfo {volume} {84}},\
  \bibinfo {pages} {066201} (\bibinfo {year} {2021})},\ \Eprint
  {http://arxiv.org/abs/2008.12090} {arXiv:2008.12090 [hep-ph]} \BibitemShut
  {NoStop}%
\bibitem [{\citenamefont {Cohen}\ \emph {et~al.}(1991)\citenamefont {Cohen},
  \citenamefont {Kaplan},\ and\ \citenamefont {Nelson}}]{Cohen:1990it}%
  \BibitemOpen
  \bibfield  {author} {\bibinfo {author} {\bibfnamefont {A.~G.}\ \bibnamefont
  {Cohen}}, \bibinfo {author} {\bibfnamefont {D.~B.}\ \bibnamefont {Kaplan}}, \
  and\ \bibinfo {author} {\bibfnamefont {A.~E.}\ \bibnamefont {Nelson}},\
  }\href {\doibase 10.1016/0550-3213(91)90395-E} {\bibfield  {journal}
  {\bibinfo  {journal} {Nucl. Phys. B}\ }\textbf {\bibinfo {volume} {349}},\
  \bibinfo {pages} {727} (\bibinfo {year} {1991})}\BibitemShut {NoStop}%
\bibitem [{\citenamefont {Cirigliano}\ \emph {et~al.}(2006)\citenamefont
  {Cirigliano}, \citenamefont {Profumo},\ and\ \citenamefont
  {Ramsey-Musolf}}]{Cirigliano:2006dg}%
  \BibitemOpen
  \bibfield  {author} {\bibinfo {author} {\bibfnamefont {V.}~\bibnamefont
  {Cirigliano}}, \bibinfo {author} {\bibfnamefont {S.}~\bibnamefont {Profumo}},
  \ and\ \bibinfo {author} {\bibfnamefont {M.~J.}\ \bibnamefont
  {Ramsey-Musolf}},\ }\href {\doibase 10.1088/1126-6708/2006/07/002} {\bibfield
   {journal} {\bibinfo  {journal} {JHEP}\ }\textbf {\bibinfo {volume} {07}},\
  \bibinfo {pages} {002} (\bibinfo {year} {2006})},\ \Eprint
  {http://arxiv.org/abs/hep-ph/0603246} {arXiv:hep-ph/0603246} \BibitemShut
  {NoStop}%
\bibitem [{\citenamefont {Morrissey}\ and\ \citenamefont
  {Ramsey-Musolf}(2012)}]{Morrissey:2012db}%
  \BibitemOpen
  \bibfield  {author} {\bibinfo {author} {\bibfnamefont {D.~E.}\ \bibnamefont
  {Morrissey}}\ and\ \bibinfo {author} {\bibfnamefont {M.~J.}\ \bibnamefont
  {Ramsey-Musolf}},\ }\href {\doibase 10.1088/1367-2630/14/12/125003}
  {\bibfield  {journal} {\bibinfo  {journal} {New J. Phys.}\ }\textbf {\bibinfo
  {volume} {14}},\ \bibinfo {pages} {125003} (\bibinfo {year} {2012})},\
  \Eprint {http://arxiv.org/abs/1206.2942} {arXiv:1206.2942 [hep-ph]}
  \BibitemShut {NoStop}%
\bibitem [{\citenamefont {Cui}\ \emph {et~al.}(2012)\citenamefont {Cui},
  \citenamefont {Randall},\ and\ \citenamefont {Shuve}}]{Cui:2011ab}%
  \BibitemOpen
  \bibfield  {author} {\bibinfo {author} {\bibfnamefont {Y.}~\bibnamefont
  {Cui}}, \bibinfo {author} {\bibfnamefont {L.}~\bibnamefont {Randall}}, \ and\
  \bibinfo {author} {\bibfnamefont {B.}~\bibnamefont {Shuve}},\ }\href
  {\doibase 10.1007/JHEP04(2012)075} {\bibfield  {journal} {\bibinfo  {journal}
  {JHEP}\ }\textbf {\bibinfo {volume} {04}},\ \bibinfo {pages} {075} (\bibinfo
  {year} {2012})},\ \Eprint {http://arxiv.org/abs/1112.2704} {arXiv:1112.2704
  [hep-ph]} \BibitemShut {NoStop}%
\bibitem [{\citenamefont {Cui}\ and\ \citenamefont
  {Sundrum}(2013)}]{Cui:2012jh}%
  \BibitemOpen
  \bibfield  {author} {\bibinfo {author} {\bibfnamefont {Y.}~\bibnamefont
  {Cui}}\ and\ \bibinfo {author} {\bibfnamefont {R.}~\bibnamefont {Sundrum}},\
  }\href {\doibase 10.1103/PhysRevD.87.116013} {\bibfield  {journal} {\bibinfo
  {journal} {Phys. Rev. D}\ }\textbf {\bibinfo {volume} {87}},\ \bibinfo
  {pages} {116013} (\bibinfo {year} {2013})},\ \Eprint
  {http://arxiv.org/abs/1212.2973} {arXiv:1212.2973 [hep-ph]} \BibitemShut
  {NoStop}%
\bibitem [{\citenamefont {Cui}\ and\ \citenamefont
  {Shuve}(2015)}]{Cui:2014twa}%
  \BibitemOpen
  \bibfield  {author} {\bibinfo {author} {\bibfnamefont {Y.}~\bibnamefont
  {Cui}}\ and\ \bibinfo {author} {\bibfnamefont {B.}~\bibnamefont {Shuve}},\
  }\href {\doibase 10.1007/JHEP02(2015)049} {\bibfield  {journal} {\bibinfo
  {journal} {JHEP}\ }\textbf {\bibinfo {volume} {02}},\ \bibinfo {pages} {049}
  (\bibinfo {year} {2015})},\ \Eprint {http://arxiv.org/abs/1409.6729}
  {arXiv:1409.6729 [hep-ph]} \BibitemShut {NoStop}%
\bibitem [{\citenamefont {Nelson}\ and\ \citenamefont
  {Xiao}(2019)}]{Nelson:2019fln}%
  \BibitemOpen
  \bibfield  {author} {\bibinfo {author} {\bibfnamefont {A.~E.}\ \bibnamefont
  {Nelson}}\ and\ \bibinfo {author} {\bibfnamefont {H.}~\bibnamefont {Xiao}},\
  }\href {\doibase 10.1103/PhysRevD.100.075002} {\bibfield  {journal} {\bibinfo
   {journal} {Phys. Rev. D}\ }\textbf {\bibinfo {volume} {100}},\ \bibinfo
  {pages} {075002} (\bibinfo {year} {2019})},\ \Eprint
  {http://arxiv.org/abs/1901.08141} {arXiv:1901.08141 [hep-ph]} \BibitemShut
  {NoStop}%
\bibitem [{\citenamefont {Cui}\ and\ \citenamefont
  {Shamma}(2020)}]{Cui:2020dly}%
  \BibitemOpen
  \bibfield  {author} {\bibinfo {author} {\bibfnamefont {Y.}~\bibnamefont
  {Cui}}\ and\ \bibinfo {author} {\bibfnamefont {M.}~\bibnamefont {Shamma}},\
  }\href {\doibase 10.1007/JHEP12(2020)046} {\bibfield  {journal} {\bibinfo
  {journal} {JHEP}\ }\textbf {\bibinfo {volume} {12}},\ \bibinfo {pages} {046}
  (\bibinfo {year} {2020})},\ \Eprint {http://arxiv.org/abs/2002.05170}
  {arXiv:2002.05170 [hep-ph]} \BibitemShut {NoStop}%
\bibitem [{\citenamefont {Beniwal}\ \emph {et~al.}(2017)\citenamefont
  {Beniwal}, \citenamefont {Lewicki}, \citenamefont {Wells}, \citenamefont
  {White},\ and\ \citenamefont {Williams}}]{Beniwal:2017eik}%
  \BibitemOpen
  \bibfield  {author} {\bibinfo {author} {\bibfnamefont {A.}~\bibnamefont
  {Beniwal}}, \bibinfo {author} {\bibfnamefont {M.}~\bibnamefont {Lewicki}},
  \bibinfo {author} {\bibfnamefont {J.~D.}\ \bibnamefont {Wells}}, \bibinfo
  {author} {\bibfnamefont {M.}~\bibnamefont {White}}, \ and\ \bibinfo {author}
  {\bibfnamefont {A.~G.}\ \bibnamefont {Williams}},\ }\href {\doibase
  10.1007/JHEP08(2017)108} {\bibfield  {journal} {\bibinfo  {journal} {JHEP}\
  }\textbf {\bibinfo {volume} {08}},\ \bibinfo {pages} {108} (\bibinfo {year}
  {2017})},\ \Eprint {http://arxiv.org/abs/1702.06124} {arXiv:1702.06124
  [hep-ph]} \BibitemShut {NoStop}%
\bibitem [{\citenamefont {Chala}\ \emph {et~al.}(2016)\citenamefont {Chala},
  \citenamefont {Nardini},\ and\ \citenamefont {Sobolev}}]{Chala:2016ykx}%
  \BibitemOpen
  \bibfield  {author} {\bibinfo {author} {\bibfnamefont {M.}~\bibnamefont
  {Chala}}, \bibinfo {author} {\bibfnamefont {G.}~\bibnamefont {Nardini}}, \
  and\ \bibinfo {author} {\bibfnamefont {I.}~\bibnamefont {Sobolev}},\ }\href
  {\doibase 10.1103/PhysRevD.94.055006} {\bibfield  {journal} {\bibinfo
  {journal} {Phys. Rev. D}\ }\textbf {\bibinfo {volume} {94}},\ \bibinfo
  {pages} {055006} (\bibinfo {year} {2016})},\ \Eprint
  {http://arxiv.org/abs/1605.08663} {arXiv:1605.08663 [hep-ph]} \BibitemShut
  {NoStop}%
\bibitem [{\citenamefont {Cui}(2015)}]{Cui:2015eba}%
  \BibitemOpen
  \bibfield  {author} {\bibinfo {author} {\bibfnamefont {Y.}~\bibnamefont
  {Cui}},\ }\href {\doibase 10.1142/S0217732315300281} {\bibfield  {journal}
  {\bibinfo  {journal} {Mod. Phys. Lett. A}\ }\textbf {\bibinfo {volume}
  {30}},\ \bibinfo {pages} {1530028} (\bibinfo {year} {2015})},\ \Eprint
  {http://arxiv.org/abs/1510.04298} {arXiv:1510.04298 [hep-ph]} \BibitemShut
  {NoStop}%
\bibitem [{\citenamefont {Cui}\ \emph {et~al.}(2016)\citenamefont {Cui},
  \citenamefont {Okui},\ and\ \citenamefont {Yunesi}}]{Cui:2016rqt}%
  \BibitemOpen
  \bibfield  {author} {\bibinfo {author} {\bibfnamefont {Y.}~\bibnamefont
  {Cui}}, \bibinfo {author} {\bibfnamefont {T.}~\bibnamefont {Okui}}, \ and\
  \bibinfo {author} {\bibfnamefont {A.}~\bibnamefont {Yunesi}},\ }\href
  {\doibase 10.1103/PhysRevD.94.115022} {\bibfield  {journal} {\bibinfo
  {journal} {Phys. Rev. D}\ }\textbf {\bibinfo {volume} {94}},\ \bibinfo
  {pages} {115022} (\bibinfo {year} {2016})},\ \Eprint
  {http://arxiv.org/abs/1605.08736} {arXiv:1605.08736 [hep-ph]} \BibitemShut
  {NoStop}%
\bibitem [{\citenamefont {Croon}\ \emph {et~al.}(2020)\citenamefont {Croon},
  \citenamefont {Howard}, \citenamefont {Ipek},\ and\ \citenamefont
  {Tait}}]{Croon:2019ugf}%
  \BibitemOpen
  \bibfield  {author} {\bibinfo {author} {\bibfnamefont {D.}~\bibnamefont
  {Croon}}, \bibinfo {author} {\bibfnamefont {J.~N.}\ \bibnamefont {Howard}},
  \bibinfo {author} {\bibfnamefont {S.}~\bibnamefont {Ipek}}, \ and\ \bibinfo
  {author} {\bibfnamefont {T.~M.~P.}\ \bibnamefont {Tait}},\ }\href {\doibase
  10.1103/PhysRevD.101.055042} {\bibfield  {journal} {\bibinfo  {journal}
  {Phys. Rev. D}\ }\textbf {\bibinfo {volume} {101}},\ \bibinfo {pages}
  {055042} (\bibinfo {year} {2020})},\ \Eprint
  {http://arxiv.org/abs/1911.01432} {arXiv:1911.01432 [hep-ph]} \BibitemShut
  {NoStop}%
\bibitem [{\citenamefont {Co}\ and\ \citenamefont
  {Harigaya}(2020)}]{Co:2019wyp}%
  \BibitemOpen
  \bibfield  {author} {\bibinfo {author} {\bibfnamefont {R.~T.}\ \bibnamefont
  {Co}}\ and\ \bibinfo {author} {\bibfnamefont {K.}~\bibnamefont {Harigaya}},\
  }\href {\doibase 10.1103/PhysRevLett.124.111602} {\bibfield  {journal}
  {\bibinfo  {journal} {Phys. Rev. Lett.}\ }\textbf {\bibinfo {volume} {124}},\
  \bibinfo {pages} {111602} (\bibinfo {year} {2020})},\ \Eprint
  {http://arxiv.org/abs/1910.02080} {arXiv:1910.02080 [hep-ph]} \BibitemShut
  {NoStop}%
\bibitem [{\citenamefont {Elor}\ and\ \citenamefont
  {McGehee}(2021)}]{Elor:2020tkc}%
  \BibitemOpen
  \bibfield  {author} {\bibinfo {author} {\bibfnamefont {G.}~\bibnamefont
  {Elor}}\ and\ \bibinfo {author} {\bibfnamefont {R.}~\bibnamefont {McGehee}},\
  }\href {\doibase 10.1103/PhysRevD.103.035005} {\bibfield  {journal} {\bibinfo
   {journal} {Phys. Rev. D}\ }\textbf {\bibinfo {volume} {103}},\ \bibinfo
  {pages} {035005} (\bibinfo {year} {2021})},\ \Eprint
  {http://arxiv.org/abs/2011.06115} {arXiv:2011.06115 [hep-ph]} \BibitemShut
  {NoStop}%
\bibitem [{\citenamefont {Davidson}\ and\ \citenamefont
  {Ibarra}(2002)}]{Davidson:2002qv}%
  \BibitemOpen
  \bibfield  {author} {\bibinfo {author} {\bibfnamefont {S.}~\bibnamefont
  {Davidson}}\ and\ \bibinfo {author} {\bibfnamefont {A.}~\bibnamefont
  {Ibarra}},\ }\href {\doibase 10.1016/S0370-2693(02)01735-5} {\bibfield
  {journal} {\bibinfo  {journal} {Phys. Lett. B}\ }\textbf {\bibinfo {volume}
  {535}},\ \bibinfo {pages} {25} (\bibinfo {year} {2002})},\ \Eprint
  {http://arxiv.org/abs/hep-ph/0202239} {arXiv:hep-ph/0202239} \BibitemShut
  {NoStop}%
\bibitem [{\citenamefont {Arkani-Hamed}\ and\ \citenamefont
  {Maldacena}(2015)}]{Arkani-Hamed:2015bza}%
  \BibitemOpen
  \bibfield  {author} {\bibinfo {author} {\bibfnamefont {N.}~\bibnamefont
  {Arkani-Hamed}}\ and\ \bibinfo {author} {\bibfnamefont {J.}~\bibnamefont
  {Maldacena}},\ }\href@noop {} {\  (\bibinfo {year} {2015})},\ \Eprint
  {http://arxiv.org/abs/1503.08043} {arXiv:1503.08043 [hep-th]} \BibitemShut
  {NoStop}%
\bibitem [{\citenamefont {Chen}\ \emph {et~al.}(2016)\citenamefont {Chen},
  \citenamefont {Wang},\ and\ \citenamefont {Xianyu}}]{Chen:2016nrs}%
  \BibitemOpen
  \bibfield  {author} {\bibinfo {author} {\bibfnamefont {X.}~\bibnamefont
  {Chen}}, \bibinfo {author} {\bibfnamefont {Y.}~\bibnamefont {Wang}}, \ and\
  \bibinfo {author} {\bibfnamefont {Z.-Z.}\ \bibnamefont {Xianyu}},\ }\href
  {\doibase 10.1007/JHEP08(2016)051} {\bibfield  {journal} {\bibinfo  {journal}
  {JHEP}\ }\textbf {\bibinfo {volume} {08}},\ \bibinfo {pages} {051} (\bibinfo
  {year} {2016})},\ \Eprint {http://arxiv.org/abs/1604.07841} {arXiv:1604.07841
  [hep-th]} \BibitemShut {NoStop}%
\bibitem [{\citenamefont {Lee}\ \emph {et~al.}(2016)\citenamefont {Lee},
  \citenamefont {Baumann},\ and\ \citenamefont {Pimentel}}]{Lee:2016vti}%
  \BibitemOpen
  \bibfield  {author} {\bibinfo {author} {\bibfnamefont {H.}~\bibnamefont
  {Lee}}, \bibinfo {author} {\bibfnamefont {D.}~\bibnamefont {Baumann}}, \ and\
  \bibinfo {author} {\bibfnamefont {G.~L.}\ \bibnamefont {Pimentel}},\ }\href
  {\doibase 10.1007/JHEP12(2016)040} {\bibfield  {journal} {\bibinfo  {journal}
  {JHEP}\ }\textbf {\bibinfo {volume} {12}},\ \bibinfo {pages} {040} (\bibinfo
  {year} {2016})},\ \Eprint {http://arxiv.org/abs/1607.03735} {arXiv:1607.03735
  [hep-th]} \BibitemShut {NoStop}%
\bibitem [{\citenamefont {Chen}\ \emph
  {et~al.}(2017{\natexlab{a}})\citenamefont {Chen}, \citenamefont {Wang},\ and\
  \citenamefont {Xianyu}}]{Chen:2016uwp}%
  \BibitemOpen
  \bibfield  {author} {\bibinfo {author} {\bibfnamefont {X.}~\bibnamefont
  {Chen}}, \bibinfo {author} {\bibfnamefont {Y.}~\bibnamefont {Wang}}, \ and\
  \bibinfo {author} {\bibfnamefont {Z.-Z.}\ \bibnamefont {Xianyu}},\ }\href
  {\doibase 10.1103/PhysRevLett.118.261302} {\bibfield  {journal} {\bibinfo
  {journal} {Phys. Rev. Lett.}\ }\textbf {\bibinfo {volume} {118}},\ \bibinfo
  {pages} {261302} (\bibinfo {year} {2017}{\natexlab{a}})},\ \Eprint
  {http://arxiv.org/abs/1610.06597} {arXiv:1610.06597 [hep-th]} \BibitemShut
  {NoStop}%
\bibitem [{\citenamefont {Chen}\ \emph
  {et~al.}(2017{\natexlab{b}})\citenamefont {Chen}, \citenamefont {Wang},\ and\
  \citenamefont {Xianyu}}]{Chen:2016hrz}%
  \BibitemOpen
  \bibfield  {author} {\bibinfo {author} {\bibfnamefont {X.}~\bibnamefont
  {Chen}}, \bibinfo {author} {\bibfnamefont {Y.}~\bibnamefont {Wang}}, \ and\
  \bibinfo {author} {\bibfnamefont {Z.-Z.}\ \bibnamefont {Xianyu}},\ }\href
  {\doibase 10.1007/JHEP04(2017)058} {\bibfield  {journal} {\bibinfo  {journal}
  {JHEP}\ }\textbf {\bibinfo {volume} {04}},\ \bibinfo {pages} {058} (\bibinfo
  {year} {2017}{\natexlab{b}})},\ \Eprint {http://arxiv.org/abs/1612.08122}
  {arXiv:1612.08122 [hep-th]} \BibitemShut {NoStop}%
\bibitem [{\citenamefont {An}\ \emph {et~al.}(2018)\citenamefont {An},
  \citenamefont {McAneny}, \citenamefont {Ridgway},\ and\ \citenamefont
  {Wise}}]{An:2017hlx}%
  \BibitemOpen
  \bibfield  {author} {\bibinfo {author} {\bibfnamefont {H.}~\bibnamefont
  {An}}, \bibinfo {author} {\bibfnamefont {M.}~\bibnamefont {McAneny}},
  \bibinfo {author} {\bibfnamefont {A.~K.}\ \bibnamefont {Ridgway}}, \ and\
  \bibinfo {author} {\bibfnamefont {M.~B.}\ \bibnamefont {Wise}},\ }\href
  {\doibase 10.1007/JHEP06(2018)105} {\bibfield  {journal} {\bibinfo  {journal}
  {JHEP}\ }\textbf {\bibinfo {volume} {06}},\ \bibinfo {pages} {105} (\bibinfo
  {year} {2018})},\ \Eprint {http://arxiv.org/abs/1706.09971} {arXiv:1706.09971
  [hep-ph]} \BibitemShut {NoStop}%
\bibitem [{\citenamefont {Kumar}\ and\ \citenamefont
  {Sundrum}(2018)}]{Kumar:2017ecc}%
  \BibitemOpen
  \bibfield  {author} {\bibinfo {author} {\bibfnamefont {S.}~\bibnamefont
  {Kumar}}\ and\ \bibinfo {author} {\bibfnamefont {R.}~\bibnamefont
  {Sundrum}},\ }\href {\doibase 10.1007/JHEP05(2018)011} {\bibfield  {journal}
  {\bibinfo  {journal} {JHEP}\ }\textbf {\bibinfo {volume} {05}},\ \bibinfo
  {pages} {011} (\bibinfo {year} {2018})},\ \Eprint
  {http://arxiv.org/abs/1711.03988} {arXiv:1711.03988 [hep-ph]} \BibitemShut
  {NoStop}%
\bibitem [{\citenamefont {Chen}\ \emph
  {et~al.}(2017{\natexlab{c}})\citenamefont {Chen}, \citenamefont {Wang},\ and\
  \citenamefont {Xianyu}}]{Chen:2017ryl}%
  \BibitemOpen
  \bibfield  {author} {\bibinfo {author} {\bibfnamefont {X.}~\bibnamefont
  {Chen}}, \bibinfo {author} {\bibfnamefont {Y.}~\bibnamefont {Wang}}, \ and\
  \bibinfo {author} {\bibfnamefont {Z.-Z.}\ \bibnamefont {Xianyu}},\ }\href
  {\doibase 10.1088/1475-7516/2017/12/006} {\bibfield  {journal} {\bibinfo
  {journal} {JCAP}\ }\textbf {\bibinfo {volume} {1712}},\ \bibinfo {pages}
  {006} (\bibinfo {year} {2017}{\natexlab{c}})},\ \Eprint
  {http://arxiv.org/abs/1703.10166} {arXiv:1703.10166 [hep-th]} \BibitemShut
  {NoStop}%
\bibitem [{\citenamefont {Chen}\ \emph {et~al.}(2018)\citenamefont {Chen},
  \citenamefont {Wang},\ and\ \citenamefont {Xianyu}}]{Chen:2018xck}%
  \BibitemOpen
  \bibfield  {author} {\bibinfo {author} {\bibfnamefont {X.}~\bibnamefont
  {Chen}}, \bibinfo {author} {\bibfnamefont {Y.}~\bibnamefont {Wang}}, \ and\
  \bibinfo {author} {\bibfnamefont {Z.-Z.}\ \bibnamefont {Xianyu}},\ }\href
  {\doibase 10.1007/JHEP09(2018)022} {\bibfield  {journal} {\bibinfo  {journal}
  {JHEP}\ }\textbf {\bibinfo {volume} {09}},\ \bibinfo {pages} {022} (\bibinfo
  {year} {2018})},\ \Eprint {http://arxiv.org/abs/1805.02656} {arXiv:1805.02656
  [hep-ph]} \BibitemShut {NoStop}%
\bibitem [{\citenamefont {Wu}(2019)}]{Wu:2018lmx}%
  \BibitemOpen
  \bibfield  {author} {\bibinfo {author} {\bibfnamefont {Y.-P.}\ \bibnamefont
  {Wu}},\ }\href {\doibase 10.1007/JHEP04(2019)125} {\bibfield  {journal}
  {\bibinfo  {journal} {JHEP}\ }\textbf {\bibinfo {volume} {04}},\ \bibinfo
  {pages} {125} (\bibinfo {year} {2019})},\ \Eprint
  {http://arxiv.org/abs/1812.10654} {arXiv:1812.10654 [hep-ph]} \BibitemShut
  {NoStop}%
\bibitem [{\citenamefont {Li}\ \emph {et~al.}(2019)\citenamefont {Li},
  \citenamefont {Nakama}, \citenamefont {Sou}, \citenamefont {Wang},\ and\
  \citenamefont {Zhou}}]{Li:2019ves}%
  \BibitemOpen
  \bibfield  {author} {\bibinfo {author} {\bibfnamefont {L.}~\bibnamefont
  {Li}}, \bibinfo {author} {\bibfnamefont {T.}~\bibnamefont {Nakama}}, \bibinfo
  {author} {\bibfnamefont {C.~M.}\ \bibnamefont {Sou}}, \bibinfo {author}
  {\bibfnamefont {Y.}~\bibnamefont {Wang}}, \ and\ \bibinfo {author}
  {\bibfnamefont {S.}~\bibnamefont {Zhou}},\ }\href {\doibase
  10.1007/JHEP07(2019)067} {\bibfield  {journal} {\bibinfo  {journal} {JHEP}\
  }\textbf {\bibinfo {volume} {07}},\ \bibinfo {pages} {067} (\bibinfo {year}
  {2019})},\ \Eprint {http://arxiv.org/abs/1903.08842} {arXiv:1903.08842
  [astro-ph.CO]} \BibitemShut {NoStop}%
\bibitem [{\citenamefont {Lu}\ \emph {et~al.}(2019)\citenamefont {Lu},
  \citenamefont {Wang},\ and\ \citenamefont {Xianyu}}]{Lu:2019tjj}%
  \BibitemOpen
  \bibfield  {author} {\bibinfo {author} {\bibfnamefont {S.}~\bibnamefont
  {Lu}}, \bibinfo {author} {\bibfnamefont {Y.}~\bibnamefont {Wang}}, \ and\
  \bibinfo {author} {\bibfnamefont {Z.-Z.}\ \bibnamefont {Xianyu}},\
  }\href@noop {} {\  (\bibinfo {year} {2019})},\ \Eprint
  {http://arxiv.org/abs/1907.07390} {arXiv:1907.07390 [hep-th]} \BibitemShut
  {NoStop}%
\bibitem [{\citenamefont {Liu}\ \emph {et~al.}(2019)\citenamefont {Liu},
  \citenamefont {Tong}, \citenamefont {Wang},\ and\ \citenamefont
  {Xianyu}}]{Liu:2019fag}%
  \BibitemOpen
  \bibfield  {author} {\bibinfo {author} {\bibfnamefont {T.}~\bibnamefont
  {Liu}}, \bibinfo {author} {\bibfnamefont {X.}~\bibnamefont {Tong}}, \bibinfo
  {author} {\bibfnamefont {Y.}~\bibnamefont {Wang}}, \ and\ \bibinfo {author}
  {\bibfnamefont {Z.-Z.}\ \bibnamefont {Xianyu}},\ }\href@noop {} {\  (\bibinfo
  {year} {2019})},\ \Eprint {http://arxiv.org/abs/1909.01819} {arXiv:1909.01819
  [hep-ph]} \BibitemShut {NoStop}%
\bibitem [{\citenamefont {Hook}\ \emph
  {et~al.}(2019{\natexlab{a}})\citenamefont {Hook}, \citenamefont {Huang},\
  and\ \citenamefont {Racco}}]{Hook:2019zxa}%
  \BibitemOpen
  \bibfield  {author} {\bibinfo {author} {\bibfnamefont {A.}~\bibnamefont
  {Hook}}, \bibinfo {author} {\bibfnamefont {J.}~\bibnamefont {Huang}}, \ and\
  \bibinfo {author} {\bibfnamefont {D.}~\bibnamefont {Racco}},\ }\href@noop {}
  {\  (\bibinfo {year} {2019}{\natexlab{a}})},\ \Eprint
  {http://arxiv.org/abs/1907.10624} {arXiv:1907.10624 [hep-ph]} \BibitemShut
  {NoStop}%
\bibitem [{\citenamefont {Hook}\ \emph
  {et~al.}(2019{\natexlab{b}})\citenamefont {Hook}, \citenamefont {Huang},\
  and\ \citenamefont {Racco}}]{Hook:2019vcn}%
  \BibitemOpen
  \bibfield  {author} {\bibinfo {author} {\bibfnamefont {A.}~\bibnamefont
  {Hook}}, \bibinfo {author} {\bibfnamefont {J.}~\bibnamefont {Huang}}, \ and\
  \bibinfo {author} {\bibfnamefont {D.}~\bibnamefont {Racco}},\ }\href@noop {}
  {\  (\bibinfo {year} {2019}{\natexlab{b}})},\ \Eprint
  {http://arxiv.org/abs/1908.00019} {arXiv:1908.00019 [hep-ph]} \BibitemShut
  {NoStop}%
\bibitem [{\citenamefont {Kumar}\ and\ \citenamefont
  {Sundrum}(2019)}]{Kumar:2019ebj}%
  \BibitemOpen
  \bibfield  {author} {\bibinfo {author} {\bibfnamefont {S.}~\bibnamefont
  {Kumar}}\ and\ \bibinfo {author} {\bibfnamefont {R.}~\bibnamefont
  {Sundrum}},\ }\href@noop {} {\  (\bibinfo {year} {2019})},\ \Eprint
  {http://arxiv.org/abs/1908.11378} {arXiv:1908.11378 [hep-ph]} \BibitemShut
  {NoStop}%
\bibitem [{\citenamefont {Alexander}\ \emph {et~al.}(2019)\citenamefont
  {Alexander}, \citenamefont {Gates}, \citenamefont {Jenks}, \citenamefont
  {Koutrolikos},\ and\ \citenamefont {McDonough}}]{Alexander:2019vtb}%
  \BibitemOpen
  \bibfield  {author} {\bibinfo {author} {\bibfnamefont {S.}~\bibnamefont
  {Alexander}}, \bibinfo {author} {\bibfnamefont {S.~J.}\ \bibnamefont
  {Gates}}, \bibinfo {author} {\bibfnamefont {L.}~\bibnamefont {Jenks}},
  \bibinfo {author} {\bibfnamefont {K.}~\bibnamefont {Koutrolikos}}, \ and\
  \bibinfo {author} {\bibfnamefont {E.}~\bibnamefont {McDonough}},\ }\href
  {\doibase 10.1007/JHEP10(2019)156} {\bibfield  {journal} {\bibinfo  {journal}
  {JHEP}\ }\textbf {\bibinfo {volume} {10}},\ \bibinfo {pages} {156} (\bibinfo
  {year} {2019})},\ \Eprint {http://arxiv.org/abs/1907.05829} {arXiv:1907.05829
  [hep-th]} \BibitemShut {NoStop}%
\bibitem [{\citenamefont {Wang}\ and\ \citenamefont
  {Xianyu}(2020{\natexlab{a}})}]{Wang:2019gbi}%
  \BibitemOpen
  \bibfield  {author} {\bibinfo {author} {\bibfnamefont {L.-T.}\ \bibnamefont
  {Wang}}\ and\ \bibinfo {author} {\bibfnamefont {Z.-Z.}\ \bibnamefont
  {Xianyu}},\ }\href {\doibase 10.1007/JHEP02(2020)044} {\bibfield  {journal}
  {\bibinfo  {journal} {JHEP}\ }\textbf {\bibinfo {volume} {02}},\ \bibinfo
  {pages} {044} (\bibinfo {year} {2020}{\natexlab{a}})},\ \Eprint
  {http://arxiv.org/abs/1910.12876} {arXiv:1910.12876 [hep-ph]} \BibitemShut
  {NoStop}%
\bibitem [{\citenamefont {Wang}(2020)}]{Wang:2019gok}%
  \BibitemOpen
  \bibfield  {author} {\bibinfo {author} {\bibfnamefont {D.-G.}\ \bibnamefont
  {Wang}},\ }\href {\doibase 10.1088/1475-7516/2020/01/046} {\bibfield
  {journal} {\bibinfo  {journal} {JCAP}\ }\textbf {\bibinfo {volume} {01}},\
  \bibinfo {pages} {046} (\bibinfo {year} {2020})},\ \Eprint
  {http://arxiv.org/abs/1911.04459} {arXiv:1911.04459 [astro-ph.CO]}
  \BibitemShut {NoStop}%
\bibitem [{\citenamefont {Wang}\ and\ \citenamefont
  {Zhu}(2020)}]{Wang:2020uic}%
  \BibitemOpen
  \bibfield  {author} {\bibinfo {author} {\bibfnamefont {Y.}~\bibnamefont
  {Wang}}\ and\ \bibinfo {author} {\bibfnamefont {Y.}~\bibnamefont {Zhu}},\
  }\href@noop {} {\  (\bibinfo {year} {2020})},\ \Eprint
  {http://arxiv.org/abs/2001.03879} {arXiv:2001.03879 [astro-ph.CO]}
  \BibitemShut {NoStop}%
\bibitem [{\citenamefont {Li}\ \emph {et~al.}(2020)\citenamefont {Li},
  \citenamefont {Lu}, \citenamefont {Wang},\ and\ \citenamefont
  {Zhou}}]{Li:2020xwr}%
  \BibitemOpen
  \bibfield  {author} {\bibinfo {author} {\bibfnamefont {L.}~\bibnamefont
  {Li}}, \bibinfo {author} {\bibfnamefont {S.}~\bibnamefont {Lu}}, \bibinfo
  {author} {\bibfnamefont {Y.}~\bibnamefont {Wang}}, \ and\ \bibinfo {author}
  {\bibfnamefont {S.}~\bibnamefont {Zhou}},\ }\href@noop {} {\  (\bibinfo
  {year} {2020})},\ \Eprint {http://arxiv.org/abs/2002.01131} {arXiv:2002.01131
  [hep-ph]} \BibitemShut {NoStop}%
\bibitem [{\citenamefont {Wang}\ and\ \citenamefont
  {Xianyu}(2020{\natexlab{b}})}]{Wang:2020ioa}%
  \BibitemOpen
  \bibfield  {author} {\bibinfo {author} {\bibfnamefont {L.-T.}\ \bibnamefont
  {Wang}}\ and\ \bibinfo {author} {\bibfnamefont {Z.-Z.}\ \bibnamefont
  {Xianyu}},\ }\href@noop {} {\  (\bibinfo {year} {2020}{\natexlab{b}})},\
  \Eprint {http://arxiv.org/abs/2004.02887} {arXiv:2004.02887 [hep-ph]}
  \BibitemShut {NoStop}%
\bibitem [{\citenamefont {Fan}\ and\ \citenamefont
  {Xianyu}(2021)}]{Fan:2020xgh}%
  \BibitemOpen
  \bibfield  {author} {\bibinfo {author} {\bibfnamefont {J.}~\bibnamefont
  {Fan}}\ and\ \bibinfo {author} {\bibfnamefont {Z.-Z.}\ \bibnamefont
  {Xianyu}},\ }\href {\doibase 10.1007/JHEP01(2021)021} {\bibfield  {journal}
  {\bibinfo  {journal} {JHEP}\ }\textbf {\bibinfo {volume} {01}},\ \bibinfo
  {pages} {021} (\bibinfo {year} {2021})},\ \Eprint
  {http://arxiv.org/abs/2005.12278} {arXiv:2005.12278 [hep-ph]} \BibitemShut
  {NoStop}%
\bibitem [{\citenamefont {Aoki}\ and\ \citenamefont
  {Yamaguchi}(2021)}]{Aoki:2020zbj}%
  \BibitemOpen
  \bibfield  {author} {\bibinfo {author} {\bibfnamefont {S.}~\bibnamefont
  {Aoki}}\ and\ \bibinfo {author} {\bibfnamefont {M.}~\bibnamefont
  {Yamaguchi}},\ }\href {\doibase 10.1007/JHEP04(2021)127} {\bibfield
  {journal} {\bibinfo  {journal} {JHEP}\ }\textbf {\bibinfo {volume} {04}},\
  \bibinfo {pages} {127} (\bibinfo {year} {2021})},\ \Eprint
  {http://arxiv.org/abs/2012.13667} {arXiv:2012.13667 [hep-th]} \BibitemShut
  {NoStop}%
\bibitem [{\citenamefont {Bodas}\ \emph {et~al.}(2021)\citenamefont {Bodas},
  \citenamefont {Kumar},\ and\ \citenamefont {Sundrum}}]{Bodas:2020yho}%
  \BibitemOpen
  \bibfield  {author} {\bibinfo {author} {\bibfnamefont {A.}~\bibnamefont
  {Bodas}}, \bibinfo {author} {\bibfnamefont {S.}~\bibnamefont {Kumar}}, \ and\
  \bibinfo {author} {\bibfnamefont {R.}~\bibnamefont {Sundrum}},\ }\href
  {\doibase 10.1007/JHEP02(2021)079} {\bibfield  {journal} {\bibinfo  {journal}
  {JHEP}\ }\textbf {\bibinfo {volume} {02}},\ \bibinfo {pages} {079} (\bibinfo
  {year} {2021})},\ \Eprint {http://arxiv.org/abs/2010.04727} {arXiv:2010.04727
  [hep-ph]} \BibitemShut {NoStop}%
\bibitem [{\citenamefont {Maru}\ and\ \citenamefont
  {Okawa}(2021)}]{Maru:2021ezc}%
  \BibitemOpen
  \bibfield  {author} {\bibinfo {author} {\bibfnamefont {N.}~\bibnamefont
  {Maru}}\ and\ \bibinfo {author} {\bibfnamefont {A.}~\bibnamefont {Okawa}},\
  }\href@noop {} {\  (\bibinfo {year} {2021})},\ \Eprint
  {http://arxiv.org/abs/2101.10634} {arXiv:2101.10634 [hep-ph]} \BibitemShut
  {NoStop}%
\bibitem [{\citenamefont {Lu}\ \emph {et~al.}(2021)\citenamefont {Lu},
  \citenamefont {Reece},\ and\ \citenamefont {Xianyu}}]{Lu:2021wxu}%
  \BibitemOpen
  \bibfield  {author} {\bibinfo {author} {\bibfnamefont {Q.}~\bibnamefont
  {Lu}}, \bibinfo {author} {\bibfnamefont {M.}~\bibnamefont {Reece}}, \ and\
  \bibinfo {author} {\bibfnamefont {Z.-Z.}\ \bibnamefont {Xianyu}},\
  }\href@noop {} {\  (\bibinfo {year} {2021})},\ \Eprint
  {http://arxiv.org/abs/2108.11385} {arXiv:2108.11385 [hep-ph]} \BibitemShut
  {NoStop}%
\bibitem [{\citenamefont {Wang}\ \emph {et~al.}(2021)\citenamefont {Wang},
  \citenamefont {Xianyu},\ and\ \citenamefont {Zhong}}]{Wang:2021qez}%
  \BibitemOpen
  \bibfield  {author} {\bibinfo {author} {\bibfnamefont {L.-T.}\ \bibnamefont
  {Wang}}, \bibinfo {author} {\bibfnamefont {Z.-Z.}\ \bibnamefont {Xianyu}}, \
  and\ \bibinfo {author} {\bibfnamefont {Y.-M.}\ \bibnamefont {Zhong}},\
  }\href@noop {} {\  (\bibinfo {year} {2021})},\ \Eprint
  {http://arxiv.org/abs/2109.14635} {arXiv:2109.14635 [hep-ph]} \BibitemShut
  {NoStop}%
\bibitem [{\citenamefont {Kim}\ \emph {et~al.}(2021)\citenamefont {Kim},
  \citenamefont {Kumar}, \citenamefont {Martin},\ and\ \citenamefont
  {Tsai}}]{Kim:2021ida}%
  \BibitemOpen
  \bibfield  {author} {\bibinfo {author} {\bibfnamefont {J.~H.}\ \bibnamefont
  {Kim}}, \bibinfo {author} {\bibfnamefont {S.}~\bibnamefont {Kumar}}, \bibinfo
  {author} {\bibfnamefont {A.}~\bibnamefont {Martin}}, \ and\ \bibinfo {author}
  {\bibfnamefont {Y.}~\bibnamefont {Tsai}},\ }\href {\doibase
  10.1007/JHEP11(2021)158} {\bibfield  {journal} {\bibinfo  {journal} {JHEP}\
  }\textbf {\bibinfo {volume} {11}},\ \bibinfo {pages} {158} (\bibinfo {year}
  {2021})},\ \Eprint {http://arxiv.org/abs/2107.09061} {arXiv:2107.09061
  [hep-ph]} \BibitemShut {NoStop}%
\bibitem [{\citenamefont {Dvali}\ \emph {et~al.}(2004)\citenamefont {Dvali},
  \citenamefont {Gruzinov},\ and\ \citenamefont {Zaldarriaga}}]{Dvali:2003em}%
  \BibitemOpen
  \bibfield  {author} {\bibinfo {author} {\bibfnamefont {G.}~\bibnamefont
  {Dvali}}, \bibinfo {author} {\bibfnamefont {A.}~\bibnamefont {Gruzinov}}, \
  and\ \bibinfo {author} {\bibfnamefont {M.}~\bibnamefont {Zaldarriaga}},\
  }\href {\doibase 10.1103/PhysRevD.69.023505} {\bibfield  {journal} {\bibinfo
  {journal} {Phys. Rev. D}\ }\textbf {\bibinfo {volume} {69}},\ \bibinfo
  {pages} {023505} (\bibinfo {year} {2004})},\ \Eprint
  {http://arxiv.org/abs/astro-ph/0303591} {arXiv:astro-ph/0303591} \BibitemShut
  {NoStop}%
\bibitem [{\citenamefont {Kofman}(2003)}]{Kofman:2003nx}%
  \BibitemOpen
  \bibfield  {author} {\bibinfo {author} {\bibfnamefont {L.}~\bibnamefont
  {Kofman}},\ }\href@noop {} {\  (\bibinfo {year} {2003})},\ \Eprint
  {http://arxiv.org/abs/astro-ph/0303614} {arXiv:astro-ph/0303614} \BibitemShut
  {NoStop}%
\bibitem [{\citenamefont {Suyama}\ and\ \citenamefont
  {Yamaguchi}(2008)}]{Suyama:2007bg}%
  \BibitemOpen
  \bibfield  {author} {\bibinfo {author} {\bibfnamefont {T.}~\bibnamefont
  {Suyama}}\ and\ \bibinfo {author} {\bibfnamefont {M.}~\bibnamefont
  {Yamaguchi}},\ }\href {\doibase 10.1103/PhysRevD.77.023505} {\bibfield
  {journal} {\bibinfo  {journal} {Phys. Rev. D}\ }\textbf {\bibinfo {volume}
  {77}},\ \bibinfo {pages} {023505} (\bibinfo {year} {2008})},\ \Eprint
  {http://arxiv.org/abs/0709.2545} {arXiv:0709.2545 [astro-ph]} \BibitemShut
  {NoStop}%
\bibitem [{\citenamefont {Ichikawa}\ \emph {et~al.}(2008)\citenamefont
  {Ichikawa}, \citenamefont {Suyama}, \citenamefont {Takahashi},\ and\
  \citenamefont {Yamaguchi}}]{Ichikawa:2008ne}%
  \BibitemOpen
  \bibfield  {author} {\bibinfo {author} {\bibfnamefont {K.}~\bibnamefont
  {Ichikawa}}, \bibinfo {author} {\bibfnamefont {T.}~\bibnamefont {Suyama}},
  \bibinfo {author} {\bibfnamefont {T.}~\bibnamefont {Takahashi}}, \ and\
  \bibinfo {author} {\bibfnamefont {M.}~\bibnamefont {Yamaguchi}},\ }\href
  {\doibase 10.1103/PhysRevD.78.063545} {\bibfield  {journal} {\bibinfo
  {journal} {Phys. Rev. D}\ }\textbf {\bibinfo {volume} {78}},\ \bibinfo
  {pages} {063545} (\bibinfo {year} {2008})},\ \Eprint
  {http://arxiv.org/abs/0807.3988} {arXiv:0807.3988 [astro-ph]} \BibitemShut
  {NoStop}%
\bibitem [{\citenamefont {Sakharov}(1967)}]{Sakharov:1967dj}%
  \BibitemOpen
  \bibfield  {author} {\bibinfo {author} {\bibfnamefont {A.~D.}\ \bibnamefont
  {Sakharov}},\ }\href {\doibase 10.1070/PU1991v034n05ABEH002497} {\bibfield
  {journal} {\bibinfo  {journal} {Pisma Zh. Eksp. Teor. Fiz.}\ }\textbf
  {\bibinfo {volume} {5}},\ \bibinfo {pages} {32} (\bibinfo {year}
  {1967})}\BibitemShut {NoStop}%
\bibitem [{\citenamefont {Tong}\ \emph {et~al.}(2021)\citenamefont {Tong},
  \citenamefont {Wang},\ and\ \citenamefont {Zhu}}]{Tong:2021wai}%
  \BibitemOpen
  \bibfield  {author} {\bibinfo {author} {\bibfnamefont {X.}~\bibnamefont
  {Tong}}, \bibinfo {author} {\bibfnamefont {Y.}~\bibnamefont {Wang}}, \ and\
  \bibinfo {author} {\bibfnamefont {Y.}~\bibnamefont {Zhu}},\ }\href@noop {} {\
   (\bibinfo {year} {2021})},\ \Eprint {http://arxiv.org/abs/2112.03448}
  {arXiv:2112.03448 [hep-th]} \BibitemShut {NoStop}%
\bibitem [{\citenamefont {Qin}\ and\ \citenamefont
  {Xianyu}(2022)}]{Qin:2022lva}%
  \BibitemOpen
  \bibfield  {author} {\bibinfo {author} {\bibfnamefont {Z.}~\bibnamefont
  {Qin}}\ and\ \bibinfo {author} {\bibfnamefont {Z.-Z.}\ \bibnamefont
  {Xianyu}},\ }\href@noop {} {\  (\bibinfo {year} {2022})},\ \Eprint
  {http://arxiv.org/abs/2205.01692} {arXiv:2205.01692 [hep-th]} \BibitemShut
  {NoStop}%
\bibitem [{\citenamefont {Chen}(2007)}]{Chen:2007fv}%
  \BibitemOpen
  \bibfield  {author} {\bibinfo {author} {\bibfnamefont {M.-C.}\ \bibnamefont
  {Chen}},\ }in\ \href@noop {} {\emph {\bibinfo {booktitle} {{Theoretical
  Advanced Study Institute in Elementary Particle Physics}: {Exploring New
  Frontiers Using Colliders and Neutrinos}}}}\ (\bibinfo {year} {2007})\ pp.\
  \bibinfo {pages} {123--176},\ \Eprint {http://arxiv.org/abs/hep-ph/0703087}
  {arXiv:hep-ph/0703087} \BibitemShut {NoStop}%
\bibitem [{\citenamefont {Kolb}\ and\ \citenamefont
  {Wolfram}(1980)}]{Kolb:1979ui}%
  \BibitemOpen
  \bibfield  {author} {\bibinfo {author} {\bibfnamefont {E.~W.}\ \bibnamefont
  {Kolb}}\ and\ \bibinfo {author} {\bibfnamefont {S.}~\bibnamefont {Wolfram}},\
  }\href {\doibase 10.1016/0370-2693(80)90435-9} {\bibfield  {journal}
  {\bibinfo  {journal} {Phys. Lett. B}\ }\textbf {\bibinfo {volume} {91}},\
  \bibinfo {pages} {217} (\bibinfo {year} {1980})}\BibitemShut {NoStop}%
\bibitem [{\citenamefont {Kolb}\ and\ \citenamefont
  {Turner}(1990)}]{Kolb:1990vq}%
  \BibitemOpen
  \bibfield  {author} {\bibinfo {author} {\bibfnamefont {E.~W.}\ \bibnamefont
  {Kolb}}\ and\ \bibinfo {author} {\bibfnamefont {M.~S.}\ \bibnamefont
  {Turner}},\ }\href@noop {} {\emph {\bibinfo {title} {{The Early
  Universe}}}},\ Vol.~\bibinfo {volume} {69}\ (\bibinfo {year}
  {1990})\BibitemShut {NoStop}%
\bibitem [{\citenamefont {Aghanim}\ \emph {et~al.}(2020)\citenamefont {Aghanim}
  \emph {et~al.}}]{Planck:2018vyg}%
  \BibitemOpen
  \bibfield  {author} {\bibinfo {author} {\bibfnamefont {N.}~\bibnamefont
  {Aghanim}} \emph {et~al.} (\bibinfo {collaboration} {Planck}),\ }\href
  {\doibase 10.1051/0004-6361/201833910} {\bibfield  {journal} {\bibinfo
  {journal} {Astron. Astrophys.}\ }\textbf {\bibinfo {volume} {641}},\ \bibinfo
  {pages} {A6} (\bibinfo {year} {2020})},\ \bibinfo {note} {[Erratum:
  Astron.Astrophys. 652, C4 (2021)]},\ \Eprint
  {http://arxiv.org/abs/1807.06209} {arXiv:1807.06209 [astro-ph.CO]}
  \BibitemShut {NoStop}%
\bibitem [{\citenamefont {Tanabashi}\ \emph {et~al.}(2018)\citenamefont
  {Tanabashi} \emph {et~al.}}]{ParticleDataGroup:2018ovx}%
  \BibitemOpen
  \bibfield  {author} {\bibinfo {author} {\bibfnamefont {M.}~\bibnamefont
  {Tanabashi}} \emph {et~al.} (\bibinfo {collaboration} {Particle Data
  Group}),\ }\href {\doibase 10.1103/PhysRevD.98.030001} {\bibfield  {journal}
  {\bibinfo  {journal} {Phys. Rev. D}\ }\textbf {\bibinfo {volume} {98}},\
  \bibinfo {pages} {030001} (\bibinfo {year} {2018})}\BibitemShut {NoStop}%
\bibitem [{\citenamefont {Meerburg}\ \emph {et~al.}(2017)\citenamefont
  {Meerburg}, \citenamefont {M{\"u}nchmeyer}, \citenamefont {Mu{\~n}oz},\ and\
  \citenamefont {Chen}}]{Meerburg:2016zdz}%
  \BibitemOpen
  \bibfield  {author} {\bibinfo {author} {\bibfnamefont {P.~D.}\ \bibnamefont
  {Meerburg}}, \bibinfo {author} {\bibfnamefont {M.}~\bibnamefont
  {M{\"u}nchmeyer}}, \bibinfo {author} {\bibfnamefont {J.~B.}\ \bibnamefont
  {Mu{\~n}oz}}, \ and\ \bibinfo {author} {\bibfnamefont {X.}~\bibnamefont
  {Chen}},\ }\href {\doibase 10.1088/1475-7516/2017/03/050} {\bibfield
  {journal} {\bibinfo  {journal} {JCAP}\ }\textbf {\bibinfo {volume} {1703}},\
  \bibinfo {pages} {050} (\bibinfo {year} {2017})},\ \Eprint
  {http://arxiv.org/abs/1610.06559} {arXiv:1610.06559 [astro-ph.CO]}
  \BibitemShut {NoStop}%
\bibitem [{\citenamefont {Moradinezhad~Dizgah}\ \emph
  {et~al.}(2018)\citenamefont {Moradinezhad~Dizgah}, \citenamefont {Lee},
  \citenamefont {Mu{\~n}oz},\ and\ \citenamefont
  {Dvorkin}}]{MoradinezhadDizgah:2018ssw}%
  \BibitemOpen
  \bibfield  {author} {\bibinfo {author} {\bibfnamefont {A.}~\bibnamefont
  {Moradinezhad~Dizgah}}, \bibinfo {author} {\bibfnamefont {H.}~\bibnamefont
  {Lee}}, \bibinfo {author} {\bibfnamefont {J.~B.}\ \bibnamefont {Mu{\~n}oz}},
  \ and\ \bibinfo {author} {\bibfnamefont {C.}~\bibnamefont {Dvorkin}},\ }\href
  {\doibase 10.1088/1475-7516/2018/05/013} {\bibfield  {journal} {\bibinfo
  {journal} {JCAP}\ }\textbf {\bibinfo {volume} {1805}},\ \bibinfo {pages}
  {013} (\bibinfo {year} {2018})},\ \Eprint {http://arxiv.org/abs/1801.07265}
  {arXiv:1801.07265 [astro-ph.CO]} \BibitemShut {NoStop}%
\bibitem [{\citenamefont {Meerburg}\ \emph {et~al.}(2019)\citenamefont
  {Meerburg} \emph {et~al.}}]{Meerburg:2019qqi}%
  \BibitemOpen
  \bibfield  {author} {\bibinfo {author} {\bibfnamefont {P.~D.}\ \bibnamefont
  {Meerburg}} \emph {et~al.},\ }\href@noop {} {\  (\bibinfo {year} {2019})},\
  \Eprint {http://arxiv.org/abs/1903.04409} {arXiv:1903.04409 [astro-ph.CO]}
  \BibitemShut {NoStop}%
\bibitem [{\citenamefont {Kogai}\ \emph {et~al.}(2021)\citenamefont {Kogai},
  \citenamefont {Akitsu}, \citenamefont {Schmidt},\ and\ \citenamefont
  {Urakawa}}]{Kogai:2020vzz}%
  \BibitemOpen
  \bibfield  {author} {\bibinfo {author} {\bibfnamefont {K.}~\bibnamefont
  {Kogai}}, \bibinfo {author} {\bibfnamefont {K.}~\bibnamefont {Akitsu}},
  \bibinfo {author} {\bibfnamefont {F.}~\bibnamefont {Schmidt}}, \ and\
  \bibinfo {author} {\bibfnamefont {Y.}~\bibnamefont {Urakawa}},\ }\href
  {\doibase 10.1088/1475-7516/2021/03/060} {\bibfield  {journal} {\bibinfo
  {journal} {JCAP}\ }\textbf {\bibinfo {volume} {03}},\ \bibinfo {pages} {060}
  (\bibinfo {year} {2021})},\ \Eprint {http://arxiv.org/abs/2009.05517}
  {arXiv:2009.05517 [astro-ph.CO]} \BibitemShut {NoStop}%
\bibitem [{\citenamefont {De~Simone}\ \emph {et~al.}(2013)\citenamefont
  {De~Simone}, \citenamefont {Perrier},\ and\ \citenamefont
  {Riotto}}]{DeSimone:2012gq}%
  \BibitemOpen
  \bibfield  {author} {\bibinfo {author} {\bibfnamefont {A.}~\bibnamefont
  {De~Simone}}, \bibinfo {author} {\bibfnamefont {H.}~\bibnamefont {Perrier}},
  \ and\ \bibinfo {author} {\bibfnamefont {A.}~\bibnamefont {Riotto}},\ }\href
  {\doibase 10.1088/1475-7516/2013/01/037} {\bibfield  {journal} {\bibinfo
  {journal} {JCAP}\ }\textbf {\bibinfo {volume} {01}},\ \bibinfo {pages} {037}
  (\bibinfo {year} {2013})},\ \Eprint {http://arxiv.org/abs/1210.6618}
  {arXiv:1210.6618 [hep-ph]} \BibitemShut {NoStop}%
\end{thebibliography}%


\clearpage
\newpage
\onecolumngrid
\setcounter{secnumdepth}{3}
\setcounter{equation}{0}
\setcounter{figure}{0}
\setcounter{table}{0}
\setcounter{page}{1}
\makeatletter
\renewcommand{\theequation}{S\arabic{equation}}
\renewcommand{\thefigure}{S\arabic{figure}}
\renewcommand{\bibnumfmt}[1]{[S#1]}
\renewcommand{\citenumfont}[1]{#1}

\begin{center}
\Large{\textbf{Probing Leptogenesis with the Cosmological Collider}}\\
\medskip
\textit{Supplementary Material}\\
\medskip
{Yanou Cui and Zhong-Zhi Xianyu}
\end{center}

\vspace{0.5cm}
In this Supplemental Material we explain details about the Higgs Yukawa couplings with three generations of Majorana neutrinos and present the full calculation for the 3-point correlator of the Higgs fluctuations for the leptogenesis model we consider.

\section{Higgs-Yukawa with 3 generations}

The discussion in the main text about the Higgs-Yukawa coupling with Majorana mass term can be directly generalized to 3 generations. The Lagrangian can be written as
\begin{align}
\label{lag3g}
  \Delta\ld=&~\nu_i^\dag\ii\ob\si^\mu\pd_\mu \nu_i+ N_i^\dag\ii\ob\si^\mu\pd_\mu N_i+\Big[m_{Dij}\big(1+\FR{h}{v}\big)\nu_i N_j-\FR{1}{2}m_{Nij} N_iN_j+\text{c.c.}\Big].
\end{align}
Here both $m_{Dij}$ and $m_{Nij}$ are complex 3$\times$3 matrices. We can use the $U(6)$ symmetry of the kinetic terms to diagonalize the full $6\times 6$ mass matrix similar as we did in the main text for 1 generation. Again we expect that the Higgs coupling matrix cannot be simultaneously diagonalized in the presence of Majorana mass, thus we have Yukawa couplings that mix different mass eigenstates and lead to unique cosmological collider signals. More explicitly, we can parameterize the $U(6)$ rotation $O$ that diagonalize the mass matrix as
\bge
   O=\bgp A & B \\ C & D \edp,
\ede
where $A, B, C, D$ are $3\times 3$ complex matrices, and $O$ satisfies the unitary condition $OO^\dag=I$. By construction,
\bge
  O^T\bgp 0 & -m_D \\  -m_D & m_N \edp O = \bgp M_1 & 0 \\ 0 & M_2 \edp,
\ede
where $M_1=\text{diag}(m_1,m_2,m_3)$ and $M_2=\text{diag}(m_4,m_5,m_6)$ represent the two diagonal blocks of the diagonalized mass matrix $M$. With this we can find the Higgs coupling matrix in the mass eigenbasis, as,
\bge
  \ld\supset \FR{h}{2v}\bgp \psi_1,\cdots,\psi_6\edp
  \bgp M_1-C^Tm_NC & -C^Tm_ND \\ -D^Tm_NC & M_2-D^Tm_ND\edp\bgp \psi_1\\ \vdots \\ \psi_6\edp,
\ede
where $\psi_i$ ($i=1,\cdots, 6$) denotes the mass eigenstate. 

It is clear that in the case of $m_N=0$, the Higgs coupling matrix is simply proportional to the mass matrix and thus the two can be simultaneously diagonalized, with non-negative real diagonal elements. No mixing could occur in the Yukawa couplings in this case. But when there is nonzero Majorana mass, the Higgs Yukawa matrix is no longer diagonalized in the mass eigenbasis. In addition, we have no further freedom to rotate $\psi_i$ as long as $m_i$ is not zero. So we would in general expect that the Yukawa matrix contains irremovable complex phases. These phases can generate the desired CP violation when the heavy neutrino decays.

\section{Diagrammatic calculations}
Here we present the full calculation using the diagrammatic rule \`a la Schwinger-Keldysh. The 3-point correlator of the Higgs fluctuation mediated by a fermion loop can be expressed as
\begin{align}
\label{3pt}
  \la\de h(\mb k_1)\de h(\mb k_2)\de h(\mb k_3)\ra'=\sum_{\mathsf{a,b}=\pm}\mathsf{ab}\int_{-\infty}^0\FR{\di\tau_1}{|H\tau_1|^4}\FR{\di\tau_2}{|H\tau_2|^4}G_\mathsf{a}(k_1,\tau_1)G_\mathsf{a}(k_2\tau_1)G_\mathsf{b}(k_3,\tau_2)\mathcal{I}_\mathsf{ab}(k_3,\tau_1,\tau_2)
\end{align} 
The function $\mathcal{I}$ is the fermionic 1-loop integral including the Yukawa couplings. According to our discussion above, there are several possibilities with distinct forms of $\mathcal{I}$. Here we show the details. First, let us review the case of a Dirac fermion, already studied in \cite{Lu:2019tjj} which can be thought of as two Weyl fermions with degenerate mass. Therefore, 
\begin{align}
  \mathcal{I}_\text{D} (k_3,\tau_1,\tau_2)=\FR{2y}{\Lambda}\int\di^3X\,e^{-\ii \mb k_s\cdot\mb X}\big\la\big[\psi\psi+\psi^\dag\psi^\dag\big](x_1)\big[\psi\psi+\psi^\dag\psi^\dag\big](x_2)\big\ra,
\end{align}
where we simply write $y_{12}=y$ and $x_1=(\tau_1,\mb X)$ and $x_2=(\tau_2,\mb 0)$ in conformal coordinates. The correlator can be worked out as
\begin{align}
  \big\la\big[\psi\psi+\psi^\dag\psi^\dag\big](x_1)\big[\psi\psi+\psi^\dag\psi^\dag\big](x_2)\big\ra=-8\Big[g_m^2(x_1,x_2)+f_m^2(x_1,x_2)\Big].
\end{align}
From \cite{Chen:2018xck},
\begin{align}
  &f_m(x,y)=-\FR{\ii H^3\Gamma(2-\ii m/H)\Gamma(2+\ii m/H)}{16\sqrt 2\pi^2}  \sqrt{1-Z}\;_{2}F_1\Big(2-\FR{\ii m}{H},2+\FR{\ii m}{H};2;\FR{1+Z}{2}\Big),\\
  &g_m(x,y)=\FR{H^3\Gamma(2-\ii m/H)\Gamma(2+\ii m/H)}{32\sqrt 2\pi^2}\FR{m}{H} \sqrt{1+Z}\;_2F_1\Big(2-\FR{\ii m}{H},2+\FR{\ii m}{H};3;\FR{1+Z}{2}\Big),
\end{align}
where $Z=Z(x,y)$ is the imbedding distance between $x$ and $y$. Let $x=(\tau_1,\mb x)$ and $y=(\tau_2,\mb y)$ in the conformal coordinates, and let $X\equiv|\mb x-\mb y|$, then $Z=1-[X^2-(\tau_1-\tau_2)^2]/(2\tau_1\tau_2)$.

To get the CHC signal at the squeezed limit, we expand this correlator at large distance limit. The result is 
\begin{align}
  -8(g_m^2+f_m^2)=-\FR{3H^6}{2\pi^5}(1-2\ii \wt m)\Gamma^2(2-\ii \wt m)\Gamma^2(-\FR{1}{2}+\ii\wt m)\Big(\FR{\tau\tau'}{x^2}\Big)^{4-2\ii \wt m}+\text{c.c.}
\end{align}
Complete the Fourier transform, we then get the nonlocal part of $\mathcal{I}$ to be 
\begin{align}
\label{diracloop}
  \mathcal{I}_\text{D}(k ,\tau_1,\tau_2)= \FR{12\ii y  H^6}{ \pi^4\Lambda k^3}(1-2\ii \wt m)\Gamma^2(2-\ii\wt m)\Gamma^2(-\FR{1}{2}+\ii\wt m)\Gamma(-6+4\ii\wt m)\sinh(2\wt m\pi)(k^2\tau_1\tau_2)^{4-2\ii\wt m}+\text{c.c.}
\end{align}

Next, consider the case of Majorana neutrino where the Higgs coupling mixes different mass eigenstates. So
\begin{align}
  \mathcal{I}_\text{M} (k_s,\tau_1,\tau_2)= \FR{y_{12}}{\Lambda}\int\di^3X\,e^{-\ii \mb k_s\cdot\mb X}\big\la\big[\psi_1\psi_2+\psi_1^\dag\psi_2^\dag\big](x_1)\big[\psi_1\psi_2+\psi_1^\dag\psi_2^\dag\big](x_2)\big\ra.
\end{align}
This time the correlator is
\begin{align}
  \big\la\big[\psi_1\psi_2+\psi_1^\dag\psi_2^\dag\big](x_1)\big[\psi_1\psi_2+\psi_1^\dag\psi_2^\dag\big](x_2)\big\ra=-4\Big[g_{m_1}(x_1,x_2)g_{m_2}(x_1,x_2)+f_{m_1}(x_1,x_2)f_{m_2}(x_1,x_2)\Big].
\end{align}
At late-time / large distance limit, we get quite different a scaling behavior,
\begin{align}
  &-4(g_{m_1}g_{m_2}+f_{m_1}f_{m_2}) = \FR{H^6}{2\pi^5}\Gamma(2+\ii\wt m_1)\Gamma(2-\ii\wt m_2)\Gamma(\FR{1}{2}-\ii\wt m_1)\Gamma(\FR{1}{2}+\ii\wt m_2)\Big(\FR{\tau_1\tau_2}{x^2}\Big)^{3+\ii(\wt m_1-\wt m_2)}+\text{c.c.}
\end{align}
That is, the correlator decreases slower than the previous case in the large $x$ limit. Fourier transform this result, we get
\begin{align}
\label{majloop}
  \mathcal{I}_\text{M}(k ,\tau_1,\tau_2)=&~\FR{-2\ii y_{12} H^6}{\pi^4\Lambda k^3}\Gamma(\FR{1}{2}-\ii\wt m_1)\Gamma(\FR{1}{2}+\ii\wt m_2)\Gamma(2+\ii\wt m_1)\Gamma(2-\ii\wt m_2)\n\\
  &~\times\Gamma(-4-2\ii\wt m_{12})\sinh(\wt m_{12}\pi)(k^2\tau_1\tau_2)^{3+\ii \wt m_{12}}+\text{c.c.},
\end{align}
where $\wt m_{12}=\wt m_1-\wt m_2$ and the tilde on any mass parameter $m$ means the dimensionless ratio $\wt m=m/H$.

Finally we consider neutrinos with nonzero CP phases. 
\begin{align}
  \mathcal{I}_\text{CP} (k_s,\tau_1,\tau_2)= |y_5y_{12}|\int\di^3x\,e^{\ii \mb k_s\cdot\mb x}\big\la\big[e^{\ii\varphi_5}\psi_1\psi_2+e^{-\ii\varphi_5}\psi_1^\dag\psi_2^\dag\big](x_1)\big[e^{\ii\varphi_{12}}\psi_1\psi_2+e^{-\ii\varphi_{12}}\psi_1^\dag\psi_2^\dag\big](x_2)\big\ra.
\end{align}
The correlator can be worked out similarly, 
\begin{align}
  & \big\la\big[e^{\ii\varphi_5}\psi_1\psi_2+e^{-\ii\varphi_5}\psi_1^\dag\psi_2^\dag\big](x_1)\big[e^{\ii\varphi_{12}}\psi_1\psi_2+e^{-\ii\varphi_{12}}\psi_1^\dag\psi_2^\dag\big](x_2)\big\ra\n\\
  =&-4 \Big[\cos(\varphi_5+\varphi_{12})g_{m_1}(x_1,x_2)g_{m_2}(x_1,x_2)+\cos(\varphi_5-\varphi_{12})f_{m_1}(x_1,x_2)f_{m_2}(x_1,x_2)\Big]\n\\
  =&~\FR{H^6}{2\pi^5}\bigg[\cos\varphi_5\cos\varphi_{12}\Gamma(2+\ii\wt m_1)\Gamma(2-\ii\wt m_2)\Gamma(\FR{1}{2}-\ii\wt m_1)\Gamma(\FR{1}{2}+\ii\wt m_2)\Big(\FR{\tau_1\tau_2}{x^2}\Big)^{3+\ii(\wt m_1-\wt m_2)}\n\\
  &~+\sin\varphi_5\sin\varphi_{12}\Gamma(2+\ii\wt m_1)\Gamma(2+\ii\wt m_2)\Gamma(\FR{1}{2}-\ii\wt m_1)\Gamma(\FR{1}{2}-\ii\wt m_2)\Big(\FR{\tau_1\tau_2}{x^2}\Big)^{3+\ii(\wt m_1+\wt m_2)}\bigg]+\text{c.c.}.
\end{align}
The result for the loop-integral in this case is:
\begin{align}
\label{cploop}
  \mathcal{I}_\text{CP}(k ,\tau_1,\tau_2)=&~\FR{-2\ii |y_{12}y_5| H^6}{\pi^4k^3}\bigg[\cos\varphi_5\cos\varphi_{12}\Gamma(\FR{1}{2}-\ii\wt m_1)\Gamma(\FR{1}{2}+\ii\wt m_2)\Gamma(2+\ii\wt m_1)\Gamma(2-\ii\wt m_2)\n\\
  &~\times\Gamma(-4-2\ii\wt m_{12})\sinh(\wt m_{12}\pi)(k^2\tau_1\tau_2)^{3+\ii \wt m_{12}}\n\\
  &~+ \sin\varphi_5\sin\varphi_{12}\Gamma(\FR{1}{2}-\ii\wt m_1)\Gamma(\FR{1}{2}-\ii\wt m_2)\Gamma(2+\ii\wt m_1)\Gamma(2+\ii\wt m_2)\n\\
  &~\times \Gamma(-4-2\ii \wt m)\sinh(\wt m\pi)(k^2\tau_1\tau_2)^{3+\ii \wt m}\bigg]+\text{c.c.}.
\end{align}

Now we are ready to calculate the 3-point correlator (\ref{3pt}) for the three cases: 1) Dirac neutrino, 2) Majorana neutrino without CP phases, 3) Majorana neutrino with CP phases. The loop integrals $\mathcal{I}$ in (\ref{3pt}) correspond to these three cases are given in (\ref{diracloop}), (\ref{majloop}), and (\ref{cploop}), respectively. 
\begin{align}
  &\mathcal{S}_\text{D}= 2\,\text{Re}\,\bigg[R_h^3 P_\zeta^{-1/2}\FR{y}{\Lambda}\mathcal{C}_\text{D}(\wt m)\Big(\FR{k_3}{2k_1}\Big)^{3-2\ii\wt m}\bigg],\\
  &\mathcal{C}_\text{D}(\wt m)=\FR{6\ii}{\pi^5}(5-2\ii\wt m)(1-\ii\wt m)^4(1-2\ii\wt m)\cosh^2(\wt m\pi)\sinh(2\wt m\pi)\n\\
  &~~~~~~~~~~~~\times\Gamma^2(1-\ii\wt m)\Gamma^2(-\FR{1}{2}+\ii m)\Gamma^2(1-2\ii\wt m)\Gamma(-6+4\ii\wt m).
\end{align}
\begin{align}
  &\mathcal{S}_\text{M}= 2\,\text{Re}\,\bigg[R_h^3 P_\zeta^{-1/2}\FR{y_{12}}{\Lambda}\mathcal{C}_\text{M}(\wt m_1,\wt m_2)\Big(\FR{k_3}{2k_1}\Big)^{2+\ii\wt m_{12}}\bigg],\\
  &\mathcal{C}_\text{M}(\wt m_1,\wt m_2)=\FR{\ii}{2\pi^5}\FR{1}{\wt m_{12}^2}(1+\FR{\ii\wt m_{12}}{4})(1-\cosh\wt m_{12}\pi)\sinh(\wt m_{12}\pi)\n\\
  &~~~~~~~~~~~\times\Gamma(\FR{1}{2}-\ii\wt m_1)\Gamma(\FR{1}{2}+\ii\wt m_2)\Gamma(2+\ii\wt m_1)\Gamma(2-\ii\wt m_2)\Gamma^2(2+\ii\wt m_{12})\Gamma(-4-2\ii\wt m_{12}).
  \label{CM}
\end{align}

\begin{align}
  &\mathcal{S}_\text{CP}= 2\,\text{Re}\,\bigg\{R_h^3 P_\zeta^{-1/2}\FR{y_{12}}{\Lambda}\bigg[\cos\varphi_5\cos\varphi_{12}\mathcal{C}_\text{M}(\wt m_1,\wt m_2)\Big(\FR{k_3}{2k_1}\Big)^{2+\ii\wt m_{12}}\n\\
  &~~~~~~~~~~~~+\sin\varphi_5\sin\varphi_{12}\mathcal{C}_\text{M}(\wt m_1,-\wt m_2)\Big(\FR{k_3}{2k_1}\Big)^{2+\ii\wt m}\bigg]\bigg\},
\end{align}
where $\mathcal{C}_\text{M}$ is given in (\ref{CM}).

\end{document}